\documentclass[aps,prl,reprint,superscriptaddress,amsmath,amssymb,bibnotes,longbibliography]{revtex4-2}

\usepackage{graphicx}
\usepackage{dcolumn}
\usepackage{bm}
\usepackage{color}
\usepackage[colorlinks,linkcolor=blue,anchorcolor=blue,citecolor=blue,urlcolor=blue,filecolor=blue,menucolor=blue,runcolor=blue]{hyperref}
\begin{document}

\title{Magnetic order and crystalline electric field excitations of the quantum critical heavy fermion ferromagnet CeRh$_6$Ge$_4$}
\author{J. W. Shu}
\affiliation{Center for Correlated Matter and Department of Physics, Zhejiang University, Hangzhou 310058, China}
\author{D. T. Adroja}
\affiliation{ISIS Facility, STFC Rutherford Appleton Laboratory, Harwell Oxford, Oxfordshire OX11 0QX, United Kingdom}
\affiliation{Highly Correlated Matter Research Group, Physics Department, University of Johannesburg, P.O. Box 524, Auckland Park 2006, South Africa}
\author{A. D. Hillier}
\affiliation{ISIS Facility, STFC Rutherford Appleton Laboratory, Harwell Oxford, Oxfordshire OX11 0QX, United Kingdom}
\author{Y. J. Zhang}
\affiliation{Institute for Advanced Materials, Hubei Normal University, Huangshi 435002, China}
\author{Y. X. Chen}
\affiliation{Center for Correlated Matter and Department of Physics, Zhejiang University, Hangzhou 310058, China}
\author{B. Shen}
\affiliation{Center for Correlated Matter and Department of Physics, Zhejiang University, Hangzhou 310058, China}
\author{F. Orlandi}
\affiliation{ISIS Facility, STFC Rutherford Appleton Laboratory, Harwell Oxford, Oxfordshire OX11 0QX, United Kingdom}
\author{H. C. Walker}
\affiliation{ISIS Facility, STFC Rutherford Appleton Laboratory, Harwell Oxford, Oxfordshire OX11 0QX, United Kingdom}
\author{Y. Liu}
\affiliation{Center for Correlated Matter and Department of Physics, Zhejiang University, Hangzhou 310058, China}
\affiliation{Zhejiang Province Key Laboratory of Quantum Technology and Device, Department of Physics, Zhejiang University, Hangzhou  310058, China}
\author{C. Cao}
\affiliation{Center for Correlated Matter and Department of Physics, Zhejiang University, Hangzhou 310058, China}
\author{F. Steglich}
\affiliation{Center for Correlated Matter and Department of Physics, Zhejiang University, Hangzhou 310058, China}
\affiliation{Max Planck Institute for Chemical Physics of Solids, Dresden, Germany}
\author{H. Q. Yuan}
\affiliation{Center for Correlated Matter and Department of Physics, Zhejiang University, Hangzhou 310058, China}
\affiliation{Zhejiang Province Key Laboratory of Quantum Technology and Device, Department of Physics, Zhejiang University, Hangzhou  310058, China}
\affiliation{State Key Laboratory of Silicon Materials, Zhejiang University, Hangzhou 310058, China}
\affiliation{Collaborative Innovation Center of Advanced Microstructures, Nanjing 210093, China}
\author{M. Smidman}
\email{msmidman@zju.edu.cn}
\affiliation{Center for Correlated Matter and Department of Physics, Zhejiang University, Hangzhou 310058, China}
\affiliation{Zhejiang Province Key Laboratory of Quantum Technology and Device, Department of Physics, Zhejiang University, Hangzhou 310058, China}

\date{\today}

\begin{abstract}
CeRh$_6$Ge$_4$ is an unusual example of a stoichiometric heavy fermion ferromagnet, which can be cleanly tuned by hydrostatic pressure to a quantum critical point. In order to understand the origin of this anomalous behavior, we have characterized the magnetic ordering and crystalline electric field (CEF) scheme of this system. While magnetic Bragg peaks are not resolved in neutron powder diffraction, coherent oscillations are observed in zero-field $\mu$SR below $T_{\rm C}$, which are consistent with in-plane ferromagnetic ordering consisting of reduced Ce moments. 
From analyzing the magnetic susceptibility and inelastic neutron scattering, we propose a CEF-level scheme which accounts for the easy-plane magnetocrystalline anisotropy, where the low lying first excited CEF exhibits significantly stronger hybridization than the ground state. These results suggest that the orbital anisotropy of the ground state and low lying  excited state doublets are important for realizing anisotropic electronic coupling between the $f$- and conduction electrons, which gives rise to the  highly anisotropic hybridization observed in photoemission experiments.
 \end{abstract}

\maketitle

In heavy fermion materials, the competition between magnetic exchange interactions which couple local moments, and the Kondo interaction between local moments and the conduction electrons, can frequently be tuned by non-thermal parameters such as pressure, magnetic fields, and/or chemical doping \cite{Weng2016}. Consequently, the antiferromagnetic (AFM) ordering temperature can often be continuously suppressed to zero at a quantum critical point (QCP), where there is a breakdown of Fermi liquid behavior, and the large accumulation of entropy can lead to emergent phases such as unconventional superconductivity  \cite{Stewart2001,Gegenwart2008}. Conversely, ferromagnetic (FM) QCP's are not usually found  \cite{Brando2016}, due to either a first-order disappearance of FM order \cite{Huxley2001,Uhlarz2004}, or the interjection of different ground states \cite{Sidorov2003,Kotegawa2013,Kaluarachchi2018}. Theoretically, it was predicted that FM QCP's are forbidden in clean itinerant FM systems \cite{Belitz1999,Chubukov2004}, and while there have been reports of their occurrence in some doped materials, including YbNi$_4$(P$_{1-x}$As$_x$)$_2$ \cite{Steppke2013}, CePd$_{0.15}$Rh$_{0.85}$ \cite{Adroja2008}, URh$_{1-x}$Ru$_x$Ge \cite{Huy2007}, and Ni$_{1-x}$Rh$_x$ \cite{Huang2020}, in such cases a disorder driven suppression of the first-order transition is difficult to exclude.

Recently, the heavy fermion ferromagnet CeRh$_6$Ge$_4$ with a Curie temperature $T_{\rm C}=2.5$~K \cite{Matsuoka2015} was found to be an exception to this paradigm, from the findings that hydrostatic  pressure continuously suppresses  $T_{\rm C}$, yielding a FM QCP \cite{Shen2019,Kotegawa2019}. The QCP is accompanied by a strange metal phase, with a linear temperature dependence of the resisitivity and a logarithmic divergence of the specific heat coefficient \cite{Shen2019}. To account for this behavior in light of the previously reported prohibition of FM QCP's in itinerant systems, it was proposed that CeRh$_6$Ge$_4$ exhibits \textit{local} quantum criticality, where the requisite entanglement of the local moments is generated by their $xy$-anisotropy \cite{Shen2019}. Moreover, in such local FM quantum critical models,  quasi-1D exchange interactions appear to be vital for avoiding a first-order transition \cite{Shen2019,Komijani2018}, which is in accordance with angle-resolved photoemission spectroscopy (ARPES) measurements revealing evidence for highly anisotropic $c-f$ coupling in CeRh$_6$Ge$_4$, from the observation of strong anisotropy in the hybridization \cite{Wu2021}. Alternatively, it was proposed that the pressure-induced first-order transition is avoided by the soft-modes, which prevent FM quantum criticality, becoming massive due to the antisymmetric spin-orbit coupling arising from the broken inversion symmetry in the crystal lattice (space group $P\bar{6}m2$) \cite{Kirkpatrick2020}. Additional studies therefore are necessary to gain insight into the origin of the FM quantum criticality, and to distinguish between the different theoretical scenarios. It is particularly important to both further characterize the nature of the magnetic ordering, and to understand the origin of the magnetic anisotropy. Therefore, we performed neutron diffraction, muon-spin relaxation ($\mu$SR), and inelastic neutron scattering measurements on polycrystalline samples of CeRh$_6$Ge$_4$.

\begin{figure} [t]
\includegraphics[angle=0,width=\columnwidth]{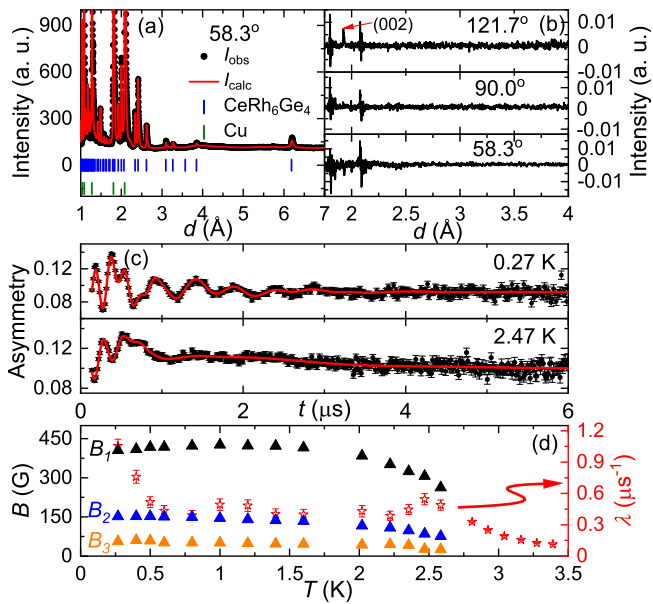}
\vspace{-10pt}
\caption{\label{Fig1} (Color online) (a) Neutron powder diffraction pattern of CeRh$_6$Ge$_4$  at 4~K, for one pair of detector banks on the WISH diffractometer. The results from a Rietveld refinement are also displayed. (b) Difference between the patterns taken at 0.27~K, and 4~K, for three pairs of banks. The scattering angles for each of the detector banks are displayed in the panels. Note that the strong features around $d=$1.8 and 2.1\AA~correspond to the (200) and (111) reflections of the Cu sample holder. (c) Zero-field $\mu$SR spectra of CeRh$_6$Ge$_4$ at two temperatures below $T_{\rm C}$, where the solid lines show the results from the fitting described in the text. (d) Temperature dependence of the internal fields (solid triangles) and Lorentzian relaxation rate $\lambda$ (open stars) obtained from fitting the $\mu$SR data.}
\end{figure}

\begin{figure*} [t]
\includegraphics[angle=0,width=0.85\textwidth]{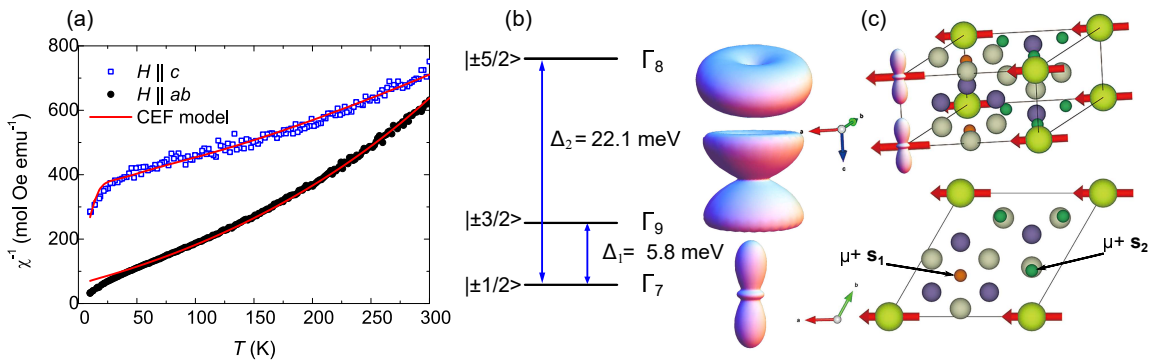}
\vspace{-10pt}
\caption{\label{Fig2} (Color online) (a) Temperature dependence of the inverse magnetic susceptibility of single crystal CeRh$_6$Ge$_4$ for two field directions \cite{Shen2019}. The solid lines show the results from fitting with a CEF model described in the text. (b) CEF level scheme and wavefunctions obtained from fitting with the CEF model, where  the angular distributions of the wave functions are also displayed. (c) Crystal structure of  CeRh$_6$Ge$_4$, with Ce, Rh, and Ge shown in yellow, grey and purple, respectively. Ferromagnetic order with moments along the $a$-axis is also illustrated, as well as the proposed muon stopping sites $\mathbf{s_1}$ and $\mathbf{s_2}$ (orange and green spheres), and the orientation of the ground state orbitals from the CEF model.}
\end{figure*}

Polycrystalline samples were prepared by arc-melting the constituent elements \cite{Matsuoka2015}, and the as-cast samples were annealed in an evacuated quartz ampoule at 1150$^\circ$C. Neutron powder diffraction measurements were performed on the WISH diffractometer at the ISIS pulsed neutron facility \cite{Chapon2011,WISHdata}, both at 4~K in the paramagnetic state, and well below $T_{\rm C}$ at 0.27~K. The data at 4~K for one pair of detector banks is shown in Fig.~\ref{Fig1}(a), where the results of a structural refinement are also displayed, with the Bragg peaks from the Cu sample holder being accounted for using the Le-Bail method, yielding lattice parameters of $a=7.144(6)$\AA~and $c=3.846(4)$\AA, in line with previous reports \cite{164growth}. In order to look for magnetic Bragg peaks, the 4~K data were subtracted from that measured at 0.27~K, which is shown for three pairs of detector banks in Fig~\ref{Fig1}(b). No additional intensity is detected at $d$-spacings corresponding to non-integer ($hkl$) reflections, in line with a lack of an AFM component to the magnetism. For FM order, the magnetic Bragg peaks are  situated on the structural peaks, and therefore weak magnetic peaks arising from small ordered moments will be more difficult to detect. No additional intensity on any nuclear peaks is consistently resolved across multiple pairs of detector banks. As shown in the top panel of  Fig~\ref{Fig1}(b), additional intensity is observed on the (002) Bragg peak at 0.27~K for one pair of banks, as expected for FM order with in-plane moments. However, this peak is not resolved for banks at different scattering angles, and no magnetic Bragg peaks at larger $d$-spacing are observed, which are expected to have greater intensity \cite{SI}. Therefore this increase might  be an artifact arising from imperfect normalization to the monitor.

Evidence for magnetic order in CeRh$_6$Ge$_4$ is however revealed by zero-field $\mu$SR measurements, which were performed on the MuSR spectrometer at the ISIS facility \cite{King2013,MUSRdata}. As shown in Fig~\ref{Fig1}(c), coherent oscillations are observed below $T_{\rm C}$, demonstrating the occurrence of long range magnetic order. The data in the magnetically ordered state are best analyzed taking into account three oscillation frequencies, using  $A(t) = A_0+\sum_{i=1}^3A_i{\rm cos}(\gamma_\mu B_it+\phi){\rm exp}[-(\sigma_it)^2/2]+A_4{\rm exp}(-\lambda t)$, while in the paramagnetic state only the first and last terms are utilized. Here $A_i$ are the initial asymmetries corresponding to local fields $B_i$, with Gaussian relaxation rates $\sigma_i$, while $\lambda$ is the Lorentzian relaxation rate of a non-oscillating component, and $\gamma_{\mu}$  is the muon-gyromagnetic ratio. The temperature dependence of the three values of $B_i$, as well as $\lambda$, are displayed in Fig.~\ref{Fig1}(d). $B_2$ and $B_3$ are found to increase with decreasing temperature, whereas $B_1$ reaches a maximum at about 1~K, and decreases slightly at lower temperatures. $\lambda$ exhibits a peak around the magnetic transition, as well as a steep increase   below about 0.8~K.

The positions of the muon stopping sites were estimated via density functional theory calculations with $f$-electrons as core electrons \cite{Wang2021,Cao2020}, from the minimum energy positions for a positive $|e|$ charge. Two crystallographically inequivalent sites were identified, $\mathbf{s_1}=(\frac{2}{3},\frac{1}{3},0)$, which corresponds to a global energy minimum, and  $\mathbf{s_2}=(0.187,0.813,0.01)$ which corresponds to a local minimum. The latter corresponds to six equivalent crystallographic positions related by three-fold and mirror symmetries, and since the three-fold symmetry is broken by in-plane FM order, this  leads to up to three distinct local fields associated with the $\mathbf{s_2}$ sites. From low temperature magnetization measurements, the ordered in-plane moment is estimated to be around $0.2-0.3\mu_{\rm B}/$Ce \cite{Shen2019}. Estimates for the local magnetic fields at the stopping sites $\mathbf{s_1}$ and  $\mathbf{s_2}$ arising from in-plane FM order were calculated using the {\sc muesr} package \cite{Bonfa2018}. For uniform FM order with a moment of $0.24\mu_{\rm B}/$Ce orientated along the $a$-axis, local fields of 364 and 152~G are calculated for the  $\mathbf{s_2}$ sites, and 87~G is calculated for  $\mathbf{s_1}$, in comparison to fitted values for $B_1$, $B_2$ and $B_3$ of 405, 151, and 56~G, respectively. On the other hand, a moment of $0.155\mu_{\rm B}/$Ce yields 58~G for  $\mathbf{s_1}$, in good agreement with $B_3$, but yields underestimates of 235 and 98~G for the $\mathbf{s_2}$ sites. In magnetic metals, an accurate comparison between the calculated and observed local fields requires accounting for the muon contact hyperfine fields, and similar discrepancies to calculations have been found for heavy fermion magnets  \cite{Gygax2004}. Moreover, a change in the hyperfine fields with temperature could lead to the non-monotonic behavior of $B_1$, which together with the increase of $\lambda$ below 0.8~K, may point to the low temperature evolution of the underlying correlated state. The differences may also arise from uncertainties in the positions of the muon stopping sites, the orientation of the moments in the basal plane,  or a spatial modulation of the ordered moment, but in the case of the latter AFM Bragg peaks would be expected to be observed in neutron diffraction.  As a result, both neutron diffraction and ZF-$\mu$SR are consistent with FM order in CeRh$_6$Ge$_4$, with a small in-plane ordered moment, and indicate the absence of any significant AFM component.

In order to determine the splitting of the $J=5/2$ Ce ground state multiplet of CeRh$_6$Ge$_4$ by  crystalline electric fields (CEF), the single crystal magnetic susceptibility \cite{Shen2019} was analyzed using the Hamiltonian  $\mathcal{H}_{\rm{CF}} = B_2^0{\rm{O_2^0}} + B_4^0{\rm{O_4^0}} $, where $B_n^m$ and $O_n^m$ are Stevens CEF parameters and operator equivalents, respectively \cite{Hutchings1964}. Note that since the Ce site has hexagonal point symmetry,  there are only two non-zero Stevens parameters  $B_2^0$ and $B_4^0$, and therefore there is no mixing of  different $|m_J\rangle$ states in the atomic wave functions. The results are displayed in Fig.~\ref{Fig2}(a), with fitted values of $B_2^0=1.25$~meV and $B_4^0=0.0056$~meV, together with molecular field parameters of $\lambda_{ab}=-52.8$~mol/emu and $\lambda_c=-111.0$~mol/emu. This yields the level scheme illustrated in Fig.~\ref{Fig2}(b), with a $\Gamma_7$ ground state doublet $\psi_{\rm GS}^{\pm}=|\pm\frac{1}{2}\rangle$, a low-lying first excited state $\psi_{1}^{\pm}=|\pm\frac{3}{2}\rangle$ separated by $\Delta_1=5.8$~meV from the ground state, and a second excited state $\psi_{2}^{\pm}=|\pm\frac{5}{2}\rangle$ at $\Delta_2=22.1$~meV. Note that a $|\pm\frac{1}{2}\rangle$ ground state is anticipated, since this is the only eigenstate which corresponds to  a non-zero in-plane moment, with $\langle \mu_x \rangle = \langle\psi^\mp|g_J(J^+ + J^-)/2|\psi^\pm\rangle=1.28\mu_{\rm B}$/Ce. Since $\langle \mu_x \rangle$ is much larger than the observed low temperature moment of around  $0.2-0.3\mu_{\rm B}/$Ce, this indicates that there is a reduced ordered moment, either due to Kondo screening processes or significant zero-point fluctuations  \cite{Adroja2003,Shen2019}. From the  large positive value of $B_2^0$, the $ab$-plane is expected to correspond to the  easy direction of magnetization, in line with the observed high and low temperature susceptibilities. This suggests that the single-ion anisotropy arising from the local environment of the Ce ions is sufficient to account for the observed easy-plane anisotropy, in contrast to many Kondo ferromagnets which  order along the hard axis \cite{Hafner2019}. While the negative values of $\lambda_{ab}$ and  $\lambda_{c}$ could indicate the presence of coexistent antiferromagnetic correlations, as in CeTi$_{1-x}$V$_x$Ge$_3$ \cite{Kittler2013,Inamdar2014,Majumder2019}, such negative values are often  found in Kondo ferromagnets \cite{Sidorov2003,Katoh2009,Krellner2012,Rai2019}, and an effective negative molecular field can arise from the Kondo effect, which scales with $T_{\rm K}$ \cite{Gruner1974,Krishna-murthy1975}.

\begin{figure} [tb]
\includegraphics[angle=0,width=\columnwidth]{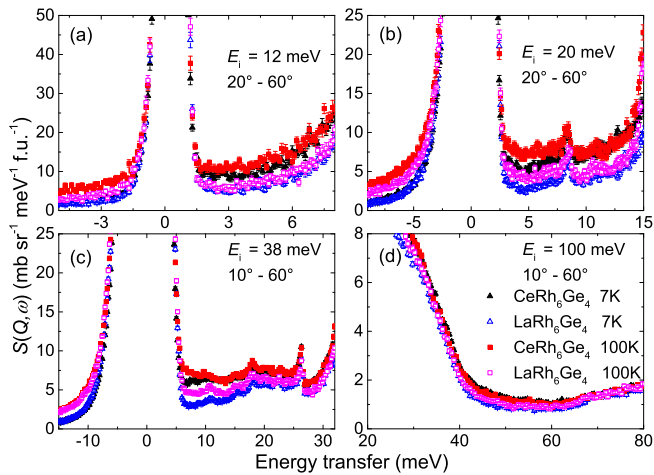}
\vspace{-10pt}
\caption{\label{Fig3} (Color online) Low angle cuts of the inelastic neutron scattering spectra of CeRh$_6$Ge$_4$ and LaRh$_6$Ge$_4$ at 7~K and 100~K for incident energies of (a) 12~meV, (b) 20~meV, (c) 38~meV, and (d) 100~meV. The integrated angular ranges are displayed in the panels.}
\end{figure}
\begin{figure} [tb]
\includegraphics[angle=0,width=0.65\columnwidth]{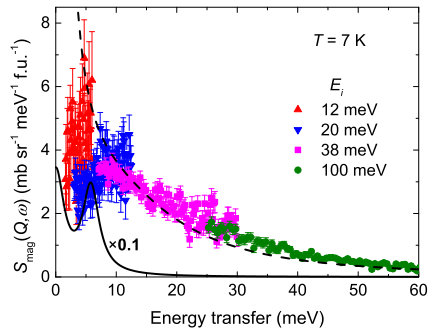}
\vspace{-10pt}
\caption{\label{Fig4} (Color online) Magnetic contribution to the inelastic neutron scattering intensity versus energy transfer at 7~K, for four different incident energies $E_i$. The solid line shows the calculated inelastic response for the CEF scheme in Fig.~\ref{Fig2}, where the FWHM of the quasielastic and inelastic peaks are 3.3~meV ($2T_K$), while the dashed line shows the case where the quasielastic FWHM is 3.3~meV but the inelastic peak corresponding to the first CEF excitation has a FWHM of 30~meV. Note that the slight mismatch between the data with  $E_i=12$ and 20~meV is an artifact arising from slight differences in the normalization for the different incident energies.}
\end{figure}

Inelastic neutron scattering measurements were performed on powder samples of CeRh$_6$Ge$_4$ and the non-magnetic analog LaRh$_6$Ge$_4$, using the MERLIN spectrometer at the ISIS facility \cite{MERLINdata}. Figure~\ref{Fig3} displays low angle cuts of the data normalized to absolute units at two temperatures, for four different incident energies $E_i$. No well defined CEF levels can be detected at energy transfers up to 80~meV. On the other hand,  over a large range of energy transfers, the scattering from CeRh$_6$Ge$_4$ is consistently larger than the La-analog indicating the presence of broad magnetic scattering, extending from at least the elastic line ($\sim1.5$~meV for $E_i=12$~meV) up to at least 60~meV. Note that magnetic scattering could not be resolved on measurements  performed at lower energies on the OSIRIS spectrometer (not displayed) \cite{OSIRISdata}, likely due to the weak and broad nature of the magnetic response. The magnetic scattering of CeRh$_6$Ge$_4$ from the MERLIN measurements was estimated by subtracting  low angle cuts of the LaRh$_6$Ge$_4$ data, taking into account the different neutron scattering cross sections, and the results are shown in  Fig.~\ref{Fig4}. It can be seen that the magnetic scattering is strongest at low energies, but with a long tail up to high energies. If this broad scattering were associated with the ground state doublet, i.e. corresponding to quasielastic scattering, this would imply a very large Kondo temperature $T_K$ on the order of hundreds of Kelvin \cite{Bauer1991}. This is in contrast to the moderate value of $T_K=19$~K deduced from comparing the magnetic entropy to a spin-1/2 Kondo model \cite{Matsuoka2015}. Due to the lack of well-defined excitations, the CEF parameters were fixed to the values from the susceptibility analysis, and the solid line in Fig.~\ref{Fig4} shows the resulting calculated inelastic neutron spectra, with a full-width at half maximum (FWHM) for all the excitations of 3.3~meV ($2T_K$). It can be seen that this fails to account for the broad magnetic scattering, and a well defined excitation at $\Delta_1$ would be expected to be observed. Note  that no excitation at $\Delta_2$ is expected, since the dipole matrix elements for the transition from $|\pm\frac{1}{2}\rangle$ to $|\pm\frac{5}{2}\rangle$ are zero due to the neutron selection rules $\Delta m_J=\pm1$. On the other hand, the dashed line shows the case with a quasielastic FWHM of $2T_K$, but a much broader inelastic excitation with a FWHM of 30~meV, and it can be seen that this scenario can well account for the broad scattering. This suggests that the inelastic neutron scattering results are consistent with the CEF scheme deduced from the magnetic susceptibility, but with the low-lying CEF excitation at 5.8~meV being significantly broadened due to hybridization with the conduction electrons.

The CEF scheme displayed in Fig.~\ref{Fig2}(b) can account for  the in-plane orientation of the ordered moments of CeRh$_6$Ge$_4$ below $T_{\rm C}$, which was proposed to be vital for generating the necessary entanglement for avoiding a first-order transition under pressure, allowing for the occurrence of a ferromagnetic QCP \cite{Shen2019}. The angular distributions of the CEF wave functions are also displayed. Notably, both the ground state and first excited doublet at 5.8~meV primarily have  electron density out of the basal plane, which may explain the strongly anisotropic hybridization revealed by  ARPES, with significantly stronger hybridization along the $c$-axis \cite{Wu2021}. Moreover, the low-lying first excited doublet appears to hybridize much more strongly with the conduction electrons, which may be a consequence of greater overlap with the out-of-plane Rh(2) and Ge(2) atoms, while the ground state charge density is orientated towards the neighboring Ce atoms [Fig.~\ref{Fig2}(c)]. Such a scenario with a more strongly hybridized excited CEF level has been predicted  to give rise to metaorbital transitions \cite{Hattori2010}, which was proposed theoretically for CeCu$_2$Si$_2$ \cite{Pourovskii2014}, yet has not been observed experimentally \cite{Amorese2020}. The influence of the first excited state on the low temperature behavior of CeRh$_6$Ge$_4$ may be inferred from the Kadowaki-Woods ratio corresponding to a ground state degeneracy $N=4$, on both sides of the QCP \cite{Shen2019}. In fact, the angular distribution of the ground state $4f$ orbitals has been identified as a key parameter for tuning the hybridization of the Ce(Co,Ir,Rh)In$_5$ family of heavy fermion superconductors \cite{Willers2010,Willers2015}, where more prolate $\Gamma_7$ ground states are associated with stronger  $c-f$ hybridization, likely due to stronger hybridization with out-of-plane In atoms \cite{Shim2007}. Our results suggest that the anisotropic hybridization is not only driven by the quasi-one-dimensional arrangement of Ce-chains, but by  the angular distribution of the CEF orbitals arising from the \textit{local} environment of the Ce atoms.

In summary, our neutron diffraction and $\mu$SR measurements are consistent with FM order in CeRh$_6$Ge$_4$, with a reduced magnetic moment compared to that expected from the CEF ground state. We propose a CEF scheme which can account for the  easy-plane anisotropy of CeRh$_6$Ge$_4$, which was predicted to be crucial for the occurrence of FM quantum criticality in this system \cite{Shen2019}. Moreover, the broad magnetic scattering observed in inelastic neutron scattering suggests the presence of strong $c-f$ hybridization, where the low lying first excited CEF level couples more strongly than the ground state. This could potentially reconcile there being significant Kondo screening processes which reduce the ordered moment, with the conclusion of  localized $4f$ electrons inferred from quantum oscillation measurements \cite{Wang2021}. These results suggest that the anisotropy of the CEF orbitals is an important factor in the observed anisotropic hybridization \cite{Wu2021}, and such anisotropic $c-f$ coupling may also give rise to quasi-1D magnetic exchange interactions, which have been proposed to avoid the first-order transition ubiquitous to isotropic systems \cite{Shen2019,Komijani2018}. As such, materials with similarly anisotropic ground state orbitals could be good candidates for searching for additional quantum critical ferromagnets.  It is of particular interest to experimentally determine if there is such a correspondingly large anisotropy in the magnetic exchange interactions of CeRh$_6$Ge$_4$, i.e. quasi-one-dimensional magnetism, which could be determined from single crystal inelastic neutron scattering or THz spectroscopy.

\begin{acknowledgements}
We are very grateful to Piers Coleman for valuable discussions, and to Franz Demmel for support with measurements on OSIRIS. This work was supported by the National Key R\&D Program of China (No.~2017YFA0303100, No.~2016YFA0300202), the National Natural Science Foundation of China (No.~12034107, No.~11874320 and No.~11974306), the Key R\&D Program of
Zhejiang Province, China (2021C01002), and  the Science Challenge Project of China (No.~TZ2016004). DTA would like to thank the Royal Society of London for Advanced Newton Fellowship funding between UK and China. Experiments at the ISIS Pulsed Neutron and Muon Source were supported by a beamtime allocation from the Science and Technology Facilities Council (RB1820492,  RB1820463,  RB1820482, RB1820611 \cite{WISHdata,MUSRdata,MERLINdata,OSIRISdata}). 
\end{acknowledgements}


\begin{thebibliography}{51}%
\makeatletter
\providecommand \@ifxundefined [1]{%
 \@ifx{#1\undefined}
}%
\providecommand \@ifnum [1]{%
 \ifnum #1\expandafter \@firstoftwo
 \else \expandafter \@secondoftwo
 \fi
}%
\providecommand \@ifx [1]{%
 \ifx #1\expandafter \@firstoftwo
 \else \expandafter \@secondoftwo
 \fi
}%
\providecommand \natexlab [1]{#1}%
\providecommand \enquote  [1]{``#1''}%
\providecommand \bibnamefont  [1]{#1}%
\providecommand \bibfnamefont [1]{#1}%
\providecommand \citenamefont [1]{#1}%
\providecommand \href@noop [0]{\@secondoftwo}%
\providecommand \href [0]{\begingroup \@sanitize@url \@href}%
\providecommand \@href[1]{\@@startlink{#1}\@@href}%
\providecommand \@@href[1]{\endgroup#1\@@endlink}%
\providecommand \@sanitize@url [0]{\catcode `\\12\catcode `\$12\catcode
  `\&12\catcode `\#12\catcode `\^12\catcode `\_12\catcode `\%12\relax}%
\providecommand \@@startlink[1]{}%
\providecommand \@@endlink[0]{}%
\providecommand \url  [0]{\begingroup\@sanitize@url \@url }%
\providecommand \@url [1]{\endgroup\@href {#1}{\urlprefix }}%
\providecommand \urlprefix  [0]{URL }%
\providecommand \Eprint [0]{\href }%
\providecommand \doibase [0]{https://doi.org/}%
\providecommand \selectlanguage [0]{\@gobble}%
\providecommand \bibinfo  [0]{\@secondoftwo}%
\providecommand \bibfield  [0]{\@secondoftwo}%
\providecommand \translation [1]{[#1]}%
\providecommand \BibitemOpen [0]{}%
\providecommand \bibitemStop [0]{}%
\providecommand \bibitemNoStop [0]{.\EOS\space}%
\providecommand \EOS [0]{\spacefactor3000\relax}%
\providecommand \BibitemShut  [1]{\csname bibitem#1\endcsname}%
\let\auto@bib@innerbib\@empty
\bibitem [{\citenamefont {Weng}\ \emph {et~al.}(2016)\citenamefont {Weng},
  \citenamefont {Smidman}, \citenamefont {Jiao}, \citenamefont {Lu},\ and\
  \citenamefont {Yuan}}]{Weng2016}%
  \BibitemOpen
  \bibfield  {author} {\bibinfo {author} {\bibfnamefont {Z.~F.}\ \bibnamefont
  {Weng}}, \bibinfo {author} {\bibfnamefont {M.}~\bibnamefont {Smidman}},
  \bibinfo {author} {\bibfnamefont {L.}~\bibnamefont {Jiao}}, \bibinfo {author}
  {\bibfnamefont {X.}~\bibnamefont {Lu}}, and\ \bibinfo {author} {\bibfnamefont
  {H.~Q.}\ \bibnamefont {Yuan}},\ }\bibfield  {title} {\bibinfo {title}
  {Multiple quantum phase transitions and superconductivity in {C}e-based heavy
  fermions},\ }\href
  {http://iopscience.iop.org/article/10.1088/0034-4885/79/9/094503/pdf}
  {\bibfield  {journal} {\bibinfo  {journal} {Reports on Progress in Physics}\
  }\textbf {\bibinfo {volume} {79}},\ \bibinfo {pages} {094503} (\bibinfo
  {year} {2016})}\BibitemShut {NoStop}%
\bibitem [{\citenamefont {Stewart}(2001)}]{Stewart2001}%
  \BibitemOpen
  \bibfield  {author} {\bibinfo {author} {\bibfnamefont {G.~R.}\ \bibnamefont
  {Stewart}},\ }\bibfield  {title} {\bibinfo {title} {Non-{F}ermi-liquid
  behavior in $d$- and $f$-electron metals},\ }\href
  {https://doi.org/10.1103/RevModPhys.73.797} {\bibfield  {journal} {\bibinfo
  {journal} {Rev. Mod. Phys.}\ }\textbf {\bibinfo {volume} {73}},\ \bibinfo
  {pages} {797} (\bibinfo {year} {2001})}\BibitemShut {NoStop}%
\bibitem [{\citenamefont {Gegenwart}\ \emph {et~al.}(2008)\citenamefont
  {Gegenwart}, \citenamefont {Si},\ and\ \citenamefont
  {Steglich}}]{Gegenwart2008}%
  \BibitemOpen
  \bibfield  {author} {\bibinfo {author} {\bibfnamefont {P.}~\bibnamefont
  {Gegenwart}}, \bibinfo {author} {\bibfnamefont {Q.}~\bibnamefont {Si}}, and\
  \bibinfo {author} {\bibfnamefont {F.}~\bibnamefont {Steglich}},\ }\bibfield
  {title} {\bibinfo {title} {Quantum criticality in heavy-fermion metals},\
  }\href {https://doi.org/10.1038/nphys892} {\bibfield  {journal} {\bibinfo
  {journal} {Nature Physics}\ }\textbf {\bibinfo {volume} {4}},\ \bibinfo
  {pages} {186} (\bibinfo {year} {2008})}\BibitemShut {NoStop}%
\bibitem [{\citenamefont {Brando}\ \emph {et~al.}(2016)\citenamefont {Brando},
  \citenamefont {Belitz}, \citenamefont {Grosche},\ and\ \citenamefont
  {Kirkpatrick}}]{Brando2016}%
  \BibitemOpen
  \bibfield  {author} {\bibinfo {author} {\bibfnamefont {M.}~\bibnamefont
  {Brando}}, \bibinfo {author} {\bibfnamefont {D.}~\bibnamefont {Belitz}},
  \bibinfo {author} {\bibfnamefont {F.~M.}\ \bibnamefont {Grosche}}, and\
  \bibinfo {author} {\bibfnamefont {T.~R.}\ \bibnamefont {Kirkpatrick}},\
  }\bibfield  {title} {\bibinfo {title} {Metallic quantum ferromagnets},\
  }\href {https://doi.org/10.1103/RevModPhys.88.025006} {\bibfield  {journal}
  {\bibinfo  {journal} {Rev. Mod. Phys.}\ }\textbf {\bibinfo {volume} {88}},\
  \bibinfo {pages} {025006} (\bibinfo {year} {2016})}\BibitemShut {NoStop}%
\bibitem [{\citenamefont {Huxley}\ \emph {et~al.}(2001)\citenamefont {Huxley},
  \citenamefont {Sheikin}, \citenamefont {Ressouche}, \citenamefont
  {Kernavanois}, \citenamefont {Braithwaite}, \citenamefont {Calemczuk},\ and\
  \citenamefont {Flouquet}}]{Huxley2001}%
  \BibitemOpen
  \bibfield  {author} {\bibinfo {author} {\bibfnamefont {A.}~\bibnamefont
  {Huxley}}, \bibinfo {author} {\bibfnamefont {I.}~\bibnamefont {Sheikin}},
  \bibinfo {author} {\bibfnamefont {E.}~\bibnamefont {Ressouche}}, \bibinfo
  {author} {\bibfnamefont {N.}~\bibnamefont {Kernavanois}}, \bibinfo {author}
  {\bibfnamefont {D.}~\bibnamefont {Braithwaite}}, \bibinfo {author}
  {\bibfnamefont {R.}~\bibnamefont {Calemczuk}}, and\ \bibinfo {author}
  {\bibfnamefont {J.}~\bibnamefont {Flouquet}},\ }\bibfield  {title} {\bibinfo
  {title} {{${\mathrm{UGe}}_{2}:$} a ferromagnetic spin-triplet
  superconductor},\ }\href {https://doi.org/10.1103/PhysRevB.63.144519}
  {\bibfield  {journal} {\bibinfo  {journal} {Phys. Rev. B}\ }\textbf {\bibinfo
  {volume} {63}},\ \bibinfo {pages} {144519} (\bibinfo {year}
  {2001})}\BibitemShut {NoStop}%
\bibitem [{\citenamefont {Uhlarz}\ \emph {et~al.}(2004)\citenamefont {Uhlarz},
  \citenamefont {Pfleiderer},\ and\ \citenamefont {Hayden}}]{Uhlarz2004}%
  \BibitemOpen
  \bibfield  {author} {\bibinfo {author} {\bibfnamefont {M.}~\bibnamefont
  {Uhlarz}}, \bibinfo {author} {\bibfnamefont {C.}~\bibnamefont {Pfleiderer}},
  and\ \bibinfo {author} {\bibfnamefont {S.~M.}\ \bibnamefont {Hayden}},\
  }\bibfield  {title} {\bibinfo {title} {Quantum phase transitions in the
  itinerant ferromagnet ${\mathrm{z}\mathrm{r}\mathrm{z}\mathrm{n}}_{2}$},\
  }\href {https://doi.org/10.1103/PhysRevLett.93.256404} {\bibfield  {journal}
  {\bibinfo  {journal} {Phys. Rev. Lett.}\ }\textbf {\bibinfo {volume} {93}},\
  \bibinfo {pages} {256404} (\bibinfo {year} {2004})}\BibitemShut {NoStop}%
\bibitem [{\citenamefont {Sidorov}\ \emph {et~al.}(2003)\citenamefont
  {Sidorov}, \citenamefont {Bauer}, \citenamefont {Frederick}, \citenamefont
  {Jeffries}, \citenamefont {Nakatsuji}, \citenamefont {Moreno}, \citenamefont
  {Thompson}, \citenamefont {Maple},\ and\ \citenamefont {Fisk}}]{Sidorov2003}%
  \BibitemOpen
  \bibfield  {author} {\bibinfo {author} {\bibfnamefont {V.~A.}\ \bibnamefont
  {Sidorov}}, \bibinfo {author} {\bibfnamefont {E.~D.}\ \bibnamefont {Bauer}},
  \bibinfo {author} {\bibfnamefont {N.~A.}\ \bibnamefont {Frederick}}, \bibinfo
  {author} {\bibfnamefont {J.~R.}\ \bibnamefont {Jeffries}}, \bibinfo {author}
  {\bibfnamefont {S.}~\bibnamefont {Nakatsuji}}, \bibinfo {author}
  {\bibfnamefont {N.~O.}\ \bibnamefont {Moreno}}, \bibinfo {author}
  {\bibfnamefont {J.~D.}\ \bibnamefont {Thompson}}, \bibinfo {author}
  {\bibfnamefont {M.~B.}\ \bibnamefont {Maple}}, and\ \bibinfo {author}
  {\bibfnamefont {Z.}~\bibnamefont {Fisk}},\ }\bibfield  {title} {\bibinfo
  {title} {Magnetic phase diagram of the ferromagnetic {Kondo-lattice compound
  ${\mathrm{CeAgSb}}_{2}$} up to 80 kbar},\ }\href
  {https://doi.org/10.1103/PhysRevB.67.224419} {\bibfield  {journal} {\bibinfo
  {journal} {Phys. Rev. B}\ }\textbf {\bibinfo {volume} {67}},\ \bibinfo
  {pages} {224419} (\bibinfo {year} {2003})}\BibitemShut {NoStop}%
\bibitem [{\citenamefont {Kotegawa}\ \emph {et~al.}(2013)\citenamefont
  {Kotegawa}, \citenamefont {Toyama}, \citenamefont {Kitagawa}, \citenamefont
  {Tou}, \citenamefont {Yamauchi}, \citenamefont {Matsuoka},\ and\
  \citenamefont {Sugawara}}]{Kotegawa2013}%
  \BibitemOpen
  \bibfield  {author} {\bibinfo {author} {\bibfnamefont {H.}~\bibnamefont
  {Kotegawa}}, \bibinfo {author} {\bibfnamefont {T.}~\bibnamefont {Toyama}},
  \bibinfo {author} {\bibfnamefont {S.}~\bibnamefont {Kitagawa}}, \bibinfo
  {author} {\bibfnamefont {H.}~\bibnamefont {Tou}}, \bibinfo {author}
  {\bibfnamefont {R.}~\bibnamefont {Yamauchi}}, \bibinfo {author}
  {\bibfnamefont {E.}~\bibnamefont {Matsuoka}}, and\ \bibinfo {author}
  {\bibfnamefont {H.}~\bibnamefont {Sugawara}},\ }\bibfield  {title} {\bibinfo
  {title} {Pressure-temperature-magnetic field phase diagram of ferromagnetic
  {K}ondo lattice {CeRuPO}},\ }\href {https://doi.org/10.7566/JPSJ.82.123711}
  {\bibfield  {journal} {\bibinfo  {journal} {J. Phys. Soc. Jpn.}\ }\textbf
  {\bibinfo {volume} {82}},\ \bibinfo {pages} {123711} (\bibinfo {year}
  {2013})}\BibitemShut {NoStop}%
\bibitem [{\citenamefont {Kaluarachchi}\ \emph {et~al.}(2018)\citenamefont
  {Kaluarachchi}, \citenamefont {Taufour}, \citenamefont {Bud'ko},\ and\
  \citenamefont {Canfield}}]{Kaluarachchi2018}%
  \BibitemOpen
  \bibfield  {author} {\bibinfo {author} {\bibfnamefont {U.~S.}\ \bibnamefont
  {Kaluarachchi}}, \bibinfo {author} {\bibfnamefont {V.}~\bibnamefont
  {Taufour}}, \bibinfo {author} {\bibfnamefont {S.~L.}\ \bibnamefont {Bud'ko}},
  and\ \bibinfo {author} {\bibfnamefont {P.~C.}\ \bibnamefont {Canfield}},\
  }\bibfield  {title} {\bibinfo {title} {Quantum tricritical point in the
  temperature-pressure-magnetic field phase diagram of
  {${\mathrm{CeTiGe}}_{3}$}},\ }\href
  {https://doi.org/10.1103/PhysRevB.97.045139} {\bibfield  {journal} {\bibinfo
  {journal} {Phys. Rev. B}\ }\textbf {\bibinfo {volume} {97}},\ \bibinfo
  {pages} {045139} (\bibinfo {year} {2018})}\BibitemShut {NoStop}%
\bibitem [{\citenamefont {Belitz}\ \emph {et~al.}(1999)\citenamefont {Belitz},
  \citenamefont {Kirkpatrick},\ and\ \citenamefont {Vojta}}]{Belitz1999}%
  \BibitemOpen
  \bibfield  {author} {\bibinfo {author} {\bibfnamefont {D.}~\bibnamefont
  {Belitz}}, \bibinfo {author} {\bibfnamefont {T.~R.}\ \bibnamefont
  {Kirkpatrick}}, and\ \bibinfo {author} {\bibfnamefont {T.}~\bibnamefont
  {Vojta}},\ }\bibfield  {title} {\bibinfo {title} {First order transitions and
  multicritical points in weak itinerant ferromagnets},\ }\href
  {https://doi.org/10.1103/PhysRevLett.82.4707} {\bibfield  {journal} {\bibinfo
   {journal} {Phys. Rev. Lett.}\ }\textbf {\bibinfo {volume} {82}},\ \bibinfo
  {pages} {4707} (\bibinfo {year} {1999})}\BibitemShut {NoStop}%
\bibitem [{\citenamefont {Chubukov}\ \emph {et~al.}(2004)\citenamefont
  {Chubukov}, \citenamefont {P\'epin},\ and\ \citenamefont
  {Rech}}]{Chubukov2004}%
  \BibitemOpen
  \bibfield  {author} {\bibinfo {author} {\bibfnamefont {A.~V.}\ \bibnamefont
  {Chubukov}}, \bibinfo {author} {\bibfnamefont {C.}~\bibnamefont {P\'epin}},
  and\ \bibinfo {author} {\bibfnamefont {J.}~\bibnamefont {Rech}},\ }\bibfield
  {title} {\bibinfo {title} {Instability of the quantum-critical point of
  itinerant ferromagnets},\ }\href
  {https://doi.org/10.1103/PhysRevLett.92.147003} {\bibfield  {journal}
  {\bibinfo  {journal} {Phys. Rev. Lett.}\ }\textbf {\bibinfo {volume} {92}},\
  \bibinfo {pages} {147003} (\bibinfo {year} {2004})}\BibitemShut {NoStop}%
\bibitem [{\citenamefont {Steppke}\ \emph {et~al.}(2013)\citenamefont
  {Steppke}, \citenamefont {K\"uchler}, \citenamefont {Lausberg}, \citenamefont
  {Lengyel}, \citenamefont {Steinke}, \citenamefont {Borth}, \citenamefont
  {L\"uhmann}, \citenamefont {Krellner}, \citenamefont {Nicklas}, \citenamefont
  {Geibel}, \citenamefont {Steglich},\ and\ \citenamefont
  {Brando}}]{Steppke2013}%
  \BibitemOpen
  \bibfield  {author} {\bibinfo {author} {\bibfnamefont {A.}~\bibnamefont
  {Steppke}}, \bibinfo {author} {\bibfnamefont {R.}~\bibnamefont {K\"uchler}},
  \bibinfo {author} {\bibfnamefont {S.}~\bibnamefont {Lausberg}}, \bibinfo
  {author} {\bibfnamefont {E.}~\bibnamefont {Lengyel}}, \bibinfo {author}
  {\bibfnamefont {L.}~\bibnamefont {Steinke}}, \bibinfo {author} {\bibfnamefont
  {R.}~\bibnamefont {Borth}}, \bibinfo {author} {\bibfnamefont
  {T.}~\bibnamefont {L\"uhmann}}, \bibinfo {author} {\bibfnamefont
  {C.}~\bibnamefont {Krellner}}, \bibinfo {author} {\bibfnamefont
  {M.}~\bibnamefont {Nicklas}}, \bibinfo {author} {\bibfnamefont
  {C.}~\bibnamefont {Geibel}}, \bibinfo {author} {\bibfnamefont
  {F.}~\bibnamefont {Steglich}}, and\ \bibinfo {author} {\bibfnamefont
  {M.}~\bibnamefont {Brando}},\ }\bibfield  {title} {\bibinfo {title}
  {Ferromagnetic quantum critical point in the heavy-fermion metal
  {$\mathrm{YbNi_4(P_{1-x}As_x)_2}$}},\ }\href
  {https://doi.org/10.1126/science.1230583} {\bibfield  {journal} {\bibinfo
  {journal} {Science}\ }\textbf {\bibinfo {volume} {339}},\ \bibinfo {pages}
  {933} (\bibinfo {year} {2013})}\BibitemShut {NoStop}%
\bibitem [{\citenamefont {Adroja}\ \emph {et~al.}(2008)\citenamefont {Adroja},
  \citenamefont {Hillier}, \citenamefont {Park}, \citenamefont {Kockelmann},
  \citenamefont {McEwen}, \citenamefont {Rainford}, \citenamefont {Jang},
  \citenamefont {Geibel},\ and\ \citenamefont {Takabatake}}]{Adroja2008}%
  \BibitemOpen
  \bibfield  {author} {\bibinfo {author} {\bibfnamefont {D.~T.}\ \bibnamefont
  {Adroja}}, \bibinfo {author} {\bibfnamefont {A.~D.}\ \bibnamefont {Hillier}},
  \bibinfo {author} {\bibfnamefont {J.-G.}\ \bibnamefont {Park}}, \bibinfo
  {author} {\bibfnamefont {W.}~\bibnamefont {Kockelmann}}, \bibinfo {author}
  {\bibfnamefont {K.~A.}\ \bibnamefont {McEwen}}, \bibinfo {author}
  {\bibfnamefont {B.~D.}\ \bibnamefont {Rainford}}, \bibinfo {author}
  {\bibfnamefont {K.-H.}\ \bibnamefont {Jang}}, \bibinfo {author}
  {\bibfnamefont {C.}~\bibnamefont {Geibel}}, and\ \bibinfo {author}
  {\bibfnamefont {T.}~\bibnamefont {Takabatake}},\ }\bibfield  {title}
  {\bibinfo {title} {Muon spin relaxation study of non-{F}ermi-liquid behavior
  near the ferromagnetic quantum critical point in
  {${\text{CePd}}_{0.15}{\text{Rh}}_{0.85}$}},\ }\href
  {https://doi.org/10.1103/PhysRevB.78.014412} {\bibfield  {journal} {\bibinfo
  {journal} {Phys. Rev. B}\ }\textbf {\bibinfo {volume} {78}},\ \bibinfo
  {pages} {014412} (\bibinfo {year} {2008})}\BibitemShut {NoStop}%
\bibitem [{\citenamefont {Huy}\ \emph {et~al.}(2007)\citenamefont {Huy},
  \citenamefont {Gasparini}, \citenamefont {Klaasse}, \citenamefont
  {de~Visser}, \citenamefont {Sakarya},\ and\ \citenamefont {van
  Dijk}}]{Huy2007}%
  \BibitemOpen
  \bibfield  {author} {\bibinfo {author} {\bibfnamefont {N.~T.}\ \bibnamefont
  {Huy}}, \bibinfo {author} {\bibfnamefont {A.}~\bibnamefont {Gasparini}},
  \bibinfo {author} {\bibfnamefont {J.~C.~P.}\ \bibnamefont {Klaasse}},
  \bibinfo {author} {\bibfnamefont {A.}~\bibnamefont {de~Visser}}, \bibinfo
  {author} {\bibfnamefont {S.}~\bibnamefont {Sakarya}}, and\ \bibinfo {author}
  {\bibfnamefont {N.~H.}\ \bibnamefont {van Dijk}},\ }\bibfield  {title}
  {\bibinfo {title} {Ferromagnetic quantum critical point in {URhGe} doped with
  {Ru}},\ }\href {https://doi.org/10.1103/PhysRevB.75.212405} {\bibfield
  {journal} {\bibinfo  {journal} {Phys. Rev. B}\ }\textbf {\bibinfo {volume}
  {75}},\ \bibinfo {pages} {212405} (\bibinfo {year} {2007})}\BibitemShut
  {NoStop}%
\bibitem [{\citenamefont {Huang}\ \emph {et~al.}(2020)\citenamefont {Huang},
  \citenamefont {Hallas}, \citenamefont {Grube}, \citenamefont {Kuntz},
  \citenamefont {Spie\ss{}}, \citenamefont {Bayliff}, \citenamefont {Besara},
  \citenamefont {Siegrist}, \citenamefont {Cai}, \citenamefont {Beare},
  \citenamefont {Luke},\ and\ \citenamefont {Morosan}}]{Huang2020}%
  \BibitemOpen
  \bibfield  {author} {\bibinfo {author} {\bibfnamefont {C.-L.}\ \bibnamefont
  {Huang}}, \bibinfo {author} {\bibfnamefont {A.~M.}\ \bibnamefont {Hallas}},
  \bibinfo {author} {\bibfnamefont {K.}~\bibnamefont {Grube}}, \bibinfo
  {author} {\bibfnamefont {S.}~\bibnamefont {Kuntz}}, \bibinfo {author}
  {\bibfnamefont {B.}~\bibnamefont {Spie\ss{}}}, \bibinfo {author}
  {\bibfnamefont {K.}~\bibnamefont {Bayliff}}, \bibinfo {author} {\bibfnamefont
  {T.}~\bibnamefont {Besara}}, \bibinfo {author} {\bibfnamefont
  {T.}~\bibnamefont {Siegrist}}, \bibinfo {author} {\bibfnamefont
  {Y.}~\bibnamefont {Cai}}, \bibinfo {author} {\bibfnamefont {J.}~\bibnamefont
  {Beare}}, \bibinfo {author} {\bibfnamefont {G.~M.}\ \bibnamefont {Luke}},
  and\ \bibinfo {author} {\bibfnamefont {E.}~\bibnamefont {Morosan}},\
  }\bibfield  {title} {\bibinfo {title} {Quantum critical point in the
  itinerant ferromagnet
  {{${\mathrm{Ni}}_{1\ensuremath{-}x}{\mathrm{Rh}}_{x}$}}},\ }\href
  {https://doi.org/10.1103/PhysRevLett.124.117203} {\bibfield  {journal}
  {\bibinfo  {journal} {Phys. Rev. Lett.}\ }\textbf {\bibinfo {volume} {124}},\
  \bibinfo {pages} {117203} (\bibinfo {year} {2020})}\BibitemShut {NoStop}%
\bibitem [{\citenamefont {Matsuoka}\ \emph {et~al.}(2015)\citenamefont
  {Matsuoka}, \citenamefont {Hondo}, \citenamefont {Fujii}, \citenamefont
  {Oshima}, \citenamefont {Sugawara}, \citenamefont {Sakurai}, \citenamefont
  {Ohta}, \citenamefont {Kneidinger}, \citenamefont {Salamakha}, \citenamefont
  {Michor},\ and\ \citenamefont {Bauer}}]{Matsuoka2015}%
  \BibitemOpen
  \bibfield  {author} {\bibinfo {author} {\bibfnamefont {E.}~\bibnamefont
  {Matsuoka}}, \bibinfo {author} {\bibfnamefont {C.}~\bibnamefont {Hondo}},
  \bibinfo {author} {\bibfnamefont {T.}~\bibnamefont {Fujii}}, \bibinfo
  {author} {\bibfnamefont {A.}~\bibnamefont {Oshima}}, \bibinfo {author}
  {\bibfnamefont {H.}~\bibnamefont {Sugawara}}, \bibinfo {author}
  {\bibfnamefont {T.}~\bibnamefont {Sakurai}}, \bibinfo {author} {\bibfnamefont
  {H.}~\bibnamefont {Ohta}}, \bibinfo {author} {\bibfnamefont {F.}~\bibnamefont
  {Kneidinger}}, \bibinfo {author} {\bibfnamefont {L.}~\bibnamefont
  {Salamakha}}, \bibinfo {author} {\bibfnamefont {H.}~\bibnamefont {Michor}},
  and\ \bibinfo {author} {\bibfnamefont {E.}~\bibnamefont {Bauer}},\ }\bibfield
   {title} {\bibinfo {title} {Ferromagnetic transition at 2.5\protect{K} in the
  hexagonal {K}ondo-lattice compound $\mathrm{CeRh_6Ge_4}$},\ }\href
  {https://doi.org/10.7566/JPSJ.84.073704} {\bibfield  {journal} {\bibinfo
  {journal} {J. Phys. Soc. Jpn.}\ }\textbf {\bibinfo {volume} {84}},\ \bibinfo
  {pages} {073704} (\bibinfo {year} {2015})}\BibitemShut {NoStop}%
\bibitem [{\citenamefont {Shen}\ \emph {et~al.}(2020)\citenamefont {Shen},
  \citenamefont {Zhang}, \citenamefont {Komijani}, \citenamefont {Nicklas},
  \citenamefont {Borth}, \citenamefont {Wang}, \citenamefont {Chen},
  \citenamefont {Nie}, \citenamefont {Li}, \citenamefont {Lu}, \citenamefont
  {Lee}, \citenamefont {Smidman}, \citenamefont {Steglich}, \citenamefont
  {Coleman},\ and\ \citenamefont {Yuan}}]{Shen2019}%
  \BibitemOpen
  \bibfield  {author} {\bibinfo {author} {\bibfnamefont {B.}~\bibnamefont
  {Shen}}, \bibinfo {author} {\bibfnamefont {Y.}~\bibnamefont {Zhang}},
  \bibinfo {author} {\bibfnamefont {Y.}~\bibnamefont {Komijani}}, \bibinfo
  {author} {\bibfnamefont {M.}~\bibnamefont {Nicklas}}, \bibinfo {author}
  {\bibfnamefont {R.}~\bibnamefont {Borth}}, \bibinfo {author} {\bibfnamefont
  {A.}~\bibnamefont {Wang}}, \bibinfo {author} {\bibfnamefont {Y.}~\bibnamefont
  {Chen}}, \bibinfo {author} {\bibfnamefont {Z.}~\bibnamefont {Nie}}, \bibinfo
  {author} {\bibfnamefont {R.}~\bibnamefont {Li}}, \bibinfo {author}
  {\bibfnamefont {X.}~\bibnamefont {Lu}}, \bibinfo {author} {\bibfnamefont
  {H.}~\bibnamefont {Lee}}, \bibinfo {author} {\bibfnamefont {M.}~\bibnamefont
  {Smidman}}, \bibinfo {author} {\bibfnamefont {F.}~\bibnamefont {Steglich}},
  \bibinfo {author} {\bibfnamefont {P.}~\bibnamefont {Coleman}}, and\ \bibinfo
  {author} {\bibfnamefont {H.}~\bibnamefont {Yuan}},\ }\bibfield  {title}
  {\bibinfo {title} {Strange metal behavior in a pure ferromagnetic {K}ondo
  lattice},\ }\href {https://doi.org/10.1038/s41586-020-2052-z} {\bibfield
  {journal} {\bibinfo  {journal} {Nature}\ }\textbf {\bibinfo {volume} {579}},\
  \bibinfo {pages} {51} (\bibinfo {year} {2020})}\BibitemShut {NoStop}%
\bibitem [{\citenamefont {Kotegawa}\ \emph {et~al.}(2019)\citenamefont
  {Kotegawa}, \citenamefont {Matsuoka}, \citenamefont {Uga}, \citenamefont
  {Takemura}, \citenamefont {Manago}, \citenamefont {Chikuchi}, \citenamefont
  {Sugawara}, \citenamefont {Tou},\ and\ \citenamefont
  {Harima}}]{Kotegawa2019}%
  \BibitemOpen
  \bibfield  {author} {\bibinfo {author} {\bibfnamefont {H.}~\bibnamefont
  {Kotegawa}}, \bibinfo {author} {\bibfnamefont {E.}~\bibnamefont {Matsuoka}},
  \bibinfo {author} {\bibfnamefont {T.}~\bibnamefont {Uga}}, \bibinfo {author}
  {\bibfnamefont {M.}~\bibnamefont {Takemura}}, \bibinfo {author}
  {\bibfnamefont {M.}~\bibnamefont {Manago}}, \bibinfo {author} {\bibfnamefont
  {N.}~\bibnamefont {Chikuchi}}, \bibinfo {author} {\bibfnamefont
  {H.}~\bibnamefont {Sugawara}}, \bibinfo {author} {\bibfnamefont
  {H.}~\bibnamefont {Tou}}, and\ \bibinfo {author} {\bibfnamefont
  {H.}~\bibnamefont {Harima}},\ }\bibfield  {title} {\bibinfo {title}
  {Indication of ferromagnetic quantum critical point in {K}ondo lattice
  {CeRh$_6$Ge$_4$}},\ }\href {https://doi.org/10.7566/JPSJ.88.093702}
  {\bibfield  {journal} {\bibinfo  {journal} {J. Phys. Soc. Jpn.}\ }\textbf
  {\bibinfo {volume} {88}},\ \bibinfo {pages} {093702} (\bibinfo {year}
  {2019})}\BibitemShut {NoStop}%
\bibitem [{\citenamefont {Komijani}\ and\ \citenamefont
  {Coleman}(2018)}]{Komijani2018}%
  \BibitemOpen
  \bibfield  {author} {\bibinfo {author} {\bibfnamefont {Y.}~\bibnamefont
  {Komijani}} and\ \bibinfo {author} {\bibfnamefont {P.}~\bibnamefont
  {Coleman}},\ }\bibfield  {title} {\bibinfo {title} {Model for a ferromagnetic
  quantum critical point in a {1D K}ondo lattice},\ }\href
  {https://doi.org/10.1103/PhysRevLett.120.157206} {\bibfield  {journal}
  {\bibinfo  {journal} {Phys. Rev. Lett.}\ }\textbf {\bibinfo {volume} {120}},\
  \bibinfo {pages} {157206} (\bibinfo {year} {2018})}\BibitemShut {NoStop}%
\bibitem [{\citenamefont {Wu}\ \emph {et~al.}(2021)\citenamefont {Wu},
  \citenamefont {Zhang}, \citenamefont {Du}, \citenamefont {Shen},
  \citenamefont {Zheng}, \citenamefont {Fang}, \citenamefont {Smidman},
  \citenamefont {Cao}, \citenamefont {Steglich}, \citenamefont {Yuan},
  \citenamefont {Denlinger},\ and\ \citenamefont {Liu}}]{Wu2021}%
  \BibitemOpen
  \bibfield  {author} {\bibinfo {author} {\bibfnamefont {Y.}~\bibnamefont
  {Wu}}, \bibinfo {author} {\bibfnamefont {Y.}~\bibnamefont {Zhang}}, \bibinfo
  {author} {\bibfnamefont {F.}~\bibnamefont {Du}}, \bibinfo {author}
  {\bibfnamefont {B.}~\bibnamefont {Shen}}, \bibinfo {author} {\bibfnamefont
  {H.}~\bibnamefont {Zheng}}, \bibinfo {author} {\bibfnamefont
  {Y.}~\bibnamefont {Fang}}, \bibinfo {author} {\bibfnamefont {M.}~\bibnamefont
  {Smidman}}, \bibinfo {author} {\bibfnamefont {C.}~\bibnamefont {Cao}},
  \bibinfo {author} {\bibfnamefont {F.}~\bibnamefont {Steglich}}, \bibinfo
  {author} {\bibfnamefont {H.}~\bibnamefont {Yuan}}, \bibinfo {author}
  {\bibfnamefont {J.~D.}\ \bibnamefont {Denlinger}}, and\ \bibinfo {author}
  {\bibfnamefont {Y.}~\bibnamefont {Liu}},\ }\bibfield  {title} {\bibinfo
  {title} {Anisotropic $c\ensuremath{-}f$ hybridization in the ferromagnetic
  quantum critical metal {${\mathrm{CeRh}}_{6}{\mathrm{Ge}}_{4}$}},\ }\href
  {https://doi.org/10.1103/PhysRevLett.126.216406} {\bibfield  {journal}
  {\bibinfo  {journal} {Phys. Rev. Lett.}\ }\textbf {\bibinfo {volume} {126}},\
  \bibinfo {pages} {216406} (\bibinfo {year} {2021})}\BibitemShut {NoStop}%
\bibitem [{\citenamefont {Kirkpatrick}\ and\ \citenamefont
  {Belitz}(2020)}]{Kirkpatrick2020}%
  \BibitemOpen
  \bibfield  {author} {\bibinfo {author} {\bibfnamefont {T.~R.}\ \bibnamefont
  {Kirkpatrick}} and\ \bibinfo {author} {\bibfnamefont {D.}~\bibnamefont
  {Belitz}},\ }\bibfield  {title} {\bibinfo {title} {Ferromagnetic quantum
  critical point in noncentrosymmetric systems},\ }\href
  {https://doi.org/10.1103/PhysRevLett.124.147201} {\bibfield  {journal}
  {\bibinfo  {journal} {Phys. Rev. Lett.}\ }\textbf {\bibinfo {volume} {124}},\
  \bibinfo {pages} {147201} (\bibinfo {year} {2020})}\BibitemShut {NoStop}%
\bibitem [{\citenamefont {Chapon}\ \emph {et~al.}(2011)\citenamefont {Chapon},
  \citenamefont {Manuel}, \citenamefont {Radaelli}, \citenamefont {Benson},
  \citenamefont {Perrott}, \citenamefont {Ansell}, \citenamefont {Rhodes},
  \citenamefont {Raspino}, \citenamefont {Duxbury}, \citenamefont {Spill},\
  and\ \citenamefont {Norris}}]{Chapon2011}%
  \BibitemOpen
  \bibfield  {author} {\bibinfo {author} {\bibfnamefont {L.~C.}\ \bibnamefont
  {Chapon}}, \bibinfo {author} {\bibfnamefont {P.}~\bibnamefont {Manuel}},
  \bibinfo {author} {\bibfnamefont {P.~G.}\ \bibnamefont {Radaelli}}, \bibinfo
  {author} {\bibfnamefont {C.}~\bibnamefont {Benson}}, \bibinfo {author}
  {\bibfnamefont {L.}~\bibnamefont {Perrott}}, \bibinfo {author} {\bibfnamefont
  {S.}~\bibnamefont {Ansell}}, \bibinfo {author} {\bibfnamefont {N.~J.}\
  \bibnamefont {Rhodes}}, \bibinfo {author} {\bibfnamefont {D.}~\bibnamefont
  {Raspino}}, \bibinfo {author} {\bibfnamefont {D.}~\bibnamefont {Duxbury}},
  \bibinfo {author} {\bibfnamefont {E.}~\bibnamefont {Spill}}, and\ \bibinfo
  {author} {\bibfnamefont {J.}~\bibnamefont {Norris}},\ }\bibfield  {title}
  {\bibinfo {title} {Wish: {T}he new powder and single crystal magnetic
  diffractometer on the second target station},\ }\href
  {https://doi.org/10.1080/10448632.2011.569650} {\bibfield  {journal}
  {\bibinfo  {journal} {Neutron News}\ }\textbf {\bibinfo {volume} {22}},\
  \bibinfo {pages} {22} (\bibinfo {year} {2011})}\BibitemShut {NoStop}%
\bibitem [{WIS()}]{WISHdata}%
  \BibitemOpen
  \href@noop {} {}\bibinfo {note} {{M. Smidman et al; (2019): STFC ISIS Neutron
  and Muon Source, \url{https://doi.org/10.5286/ISIS.E.RB1820482}}}\BibitemShut
  {NoStop}%
\bibitem [{\citenamefont {Vosswinkel}\ \emph {et~al.}(2012)\citenamefont
  {Vosswinkel}, \citenamefont {Niehaus}, \citenamefont {Rodewald},\ and\
  \citenamefont {Pottgen}}]{164growth}%
  \BibitemOpen
  \bibfield  {author} {\bibinfo {author} {\bibfnamefont {D.}~\bibnamefont
  {Vosswinkel}}, \bibinfo {author} {\bibfnamefont {O.}~\bibnamefont {Niehaus}},
  \bibinfo {author} {\bibfnamefont {U.~C.}\ \bibnamefont {Rodewald}}, and\
  \bibinfo {author} {\bibfnamefont {R.}~\bibnamefont {Pottgen}},\ }\bibfield
  {title} {\bibinfo {title} {Bismuth flux growth of $\mathrm{CeRh_6Ge_4}$ and
  $\mathrm{CeRh_2Ge_2}$ single crystals},\ }\href
  {https://doi.org/10.5560/ZNB.2012-0265} {\bibfield  {journal} {\bibinfo
  {journal} {Zeitschrift f\"ur Naturforschung B}\ }\textbf {\bibinfo {volume}
  {67}},\ \bibinfo {pages} {1241} (\bibinfo {year} {2012})}\BibitemShut
  {NoStop}%
\bibitem [{SI()}]{SI}%
  \BibitemOpen
  \href@noop {} {}\bibinfo {note} {See Supplemental Material at [] for the
  simulated magnetic diffraction pattern corresponding to in-plane
  ferromagnetic order.}\BibitemShut {Stop}%
\bibitem [{\citenamefont {King}\ \emph {et~al.}(2013)\citenamefont {King},
  \citenamefont {de~Renzi}, \citenamefont {Cottrell}, \citenamefont {Hillier},\
  and\ \citenamefont {Cox}}]{King2013}%
  \BibitemOpen
  \bibfield  {author} {\bibinfo {author} {\bibfnamefont {P.~J.~C.}\
  \bibnamefont {King}}, \bibinfo {author} {\bibfnamefont {R.}~\bibnamefont
  {de~Renzi}}, \bibinfo {author} {\bibfnamefont {S.~P.}\ \bibnamefont
  {Cottrell}}, \bibinfo {author} {\bibfnamefont {A.~D.}\ \bibnamefont
  {Hillier}}, and\ \bibinfo {author} {\bibfnamefont {S.~F.~J.}\ \bibnamefont
  {Cox}},\ }\bibfield  {title} {\bibinfo {title} {{ISIS} muons for materials
  and molecular science studies},\ }\href
  {https://doi.org/10.1088/0031-8949/88/06/068502} {\bibfield  {journal}
  {\bibinfo  {journal} {Physica Scripta}\ }\textbf {\bibinfo {volume} {88}},\
  \bibinfo {pages} {068502} (\bibinfo {year} {2013})}\BibitemShut {NoStop}%
\bibitem [{MUS()}]{MUSRdata}%
  \BibitemOpen
  \href@noop {} {}\bibinfo {note} {{M. Smidman et al; (2018): STFC ISIS Neutron
  and Muon Source, \url{https://doi.org/10.5286/ISIS.E.99690746}}}\BibitemShut
  {NoStop}%
\bibitem [{\citenamefont {Wang}\ \emph {et~al.}(2021)\citenamefont {Wang},
  \citenamefont {Du}, \citenamefont {Zhang}, \citenamefont {Graf},
  \citenamefont {Shen}, \citenamefont {Chen}, \citenamefont {Liu},
  \citenamefont {Smidman}, \citenamefont {Cao}, \citenamefont {Steglich},\ and\
  \citenamefont {Yuan}}]{Wang2021}%
  \BibitemOpen
  \bibfield  {author} {\bibinfo {author} {\bibfnamefont {A.}~\bibnamefont
  {Wang}}, \bibinfo {author} {\bibfnamefont {F.}~\bibnamefont {Du}}, \bibinfo
  {author} {\bibfnamefont {Y.~J.}~\bibnamefont {Zhang}}, \bibinfo {author}
  {\bibfnamefont {D.}~\bibnamefont {Graf}}, \bibinfo {author} {\bibfnamefont
  {B.}~\bibnamefont {Shen}}, \bibinfo {author} {\bibfnamefont {Y.}~\bibnamefont
  {Chen}}, \bibinfo {author} {\bibfnamefont {Y.}~\bibnamefont {Liu}}, \bibinfo
  {author} {\bibfnamefont {M.}~\bibnamefont {Smidman}}, \bibinfo {author}
  {\bibfnamefont {C.}~\bibnamefont {Cao}}, \bibinfo {author} {\bibfnamefont
  {F.}~\bibnamefont {Steglich}}, and\ \bibinfo {author} {\bibfnamefont
  {H.~Q.}~\bibnamefont {Yuan}},\ }\bibfield  {title} {\bibinfo {title} {Localized
  $4f$-electrons in the quantum critical heavy fermion ferromagnet
  {CeRh$_6$Ge$_4$}},\ }\href
  {https://www.sciencedirect.com/science/article/pii/S2095927321001833}
  {\bibfield  {journal} {\bibinfo  {journal} {Science Bulletin}\ }\textbf {\bibinfo {volume} {66}},\
  \bibinfo {pages} {1389} (\bibinfo
  {year} {2021})}\BibitemShut {NoStop}%
\bibitem [{\citenamefont {Cao}\ and\ \citenamefont {Zhu}(2020)}]{Cao2020}%
  \BibitemOpen
  \bibfield  {author} {\bibinfo {author} {\bibfnamefont {C.}~\bibnamefont
  {Cao}} and\ \bibinfo {author} {\bibfnamefont {J.-X.}\ \bibnamefont {Zhu}},\
  }\bibfield  {title} {\bibinfo {title} {Pressure dependent electronic
  structure in {CeRh$_6$Ge$_4$}},\ }\href {https://arxiv.org/abs/2011.14256}
  {\bibfield  {journal} {\bibinfo  {journal} {arXiv:2011.14256}\ } (\bibinfo
  {year} {2020})}\BibitemShut {NoStop}%
\bibitem [{\citenamefont {Bonf{\'a}}\ \emph {et~al.}(2018)\citenamefont
  {Bonf{\'a}}, \citenamefont {Onuorah},\ and\ \citenamefont
  {Renzi}}]{Bonfa2018}%
  \BibitemOpen
  \bibfield  {author} {\bibinfo {author} {\bibfnamefont {P.}~\bibnamefont
  {Bonf{\'a}}}, \bibinfo {author} {\bibfnamefont {I.~J.}\ \bibnamefont
  {Onuorah}}, and\ \bibinfo {author} {\bibfnamefont {R.~D.}\ \bibnamefont
  {Renzi}},\ }\bibfield  {title} {\bibinfo {title} {Introduction and a quick
  look at {MUESR, the Magnetic Structure and mUon Embedding Site Refinement
  Suite}},\ }\href {https://doi.org/10.7566/JPSCP.21.011052} {\bibfield
  {journal} {\bibinfo  {journal} {JPS Conf. Proc.}\ }\textbf {\bibinfo {volume}
  {21}},\ \bibinfo {pages} {011052} (\bibinfo {year} {2018})},\ \bibinfo {note}
  {{Proceedings of the 14th International Conference on Muon Spin Rotation,
  Relaxation and Resonance ($\mu$SR2017)}}\BibitemShut {NoStop}%
\bibitem [{\citenamefont {Gygax}\ \emph {et~al.}(2004)\citenamefont {Gygax},
  \citenamefont {Schenck},\ and\ \citenamefont {\={O}nuki}}]{Gygax2004}%
  \BibitemOpen
  \bibfield  {author} {\bibinfo {author} {\bibfnamefont {F.~N.}\ \bibnamefont
  {Gygax}}, \bibinfo {author} {\bibfnamefont {A.}~\bibnamefont {Schenck}}, and\
  \bibinfo {author} {\bibfnamefont {Y.}~\bibnamefont {\={O}nuki}},\ }\bibfield
  {title} {\bibinfo {title} {Magnetic properties of {CeCu}$_2$ tested by
  muon-spin rotation and relaxation},\ }\href
  {https://doi.org/10.1088/0953-8984/16/13/020} {\bibfield  {journal} {\bibinfo
   {journal} {J. Phys.: Condens. Matter}\ }\textbf {\bibinfo {volume} {16}},\
  \bibinfo {pages} {2421} (\bibinfo {year} {2004})}\BibitemShut {NoStop}%
\bibitem [{\citenamefont {Hutchings}(1964)}]{Hutchings1964}%
  \BibitemOpen
  \bibfield  {author} {\bibinfo {author} {\bibfnamefont {M.}~\bibnamefont
  {Hutchings}},\ }\bibfield  {title} {\bibinfo {title} {Point-charge
  calculations of energy levels of magnetic ions in crystalline electric
  fields}\ }(\bibinfo  {publisher} {Academic Press},\ \bibinfo {year} {1964})\
  pp.\ \bibinfo {pages} {227 -- 273}\BibitemShut {NoStop}%
\bibitem [{\citenamefont {Adroja}\ \emph {et~al.}(2003)\citenamefont {Adroja},
  \citenamefont {Kockelmann}, \citenamefont {Hillier}, \citenamefont {So},
  \citenamefont {Knight},\ and\ \citenamefont {Rainford}}]{Adroja2003}%
  \BibitemOpen
  \bibfield  {author} {\bibinfo {author} {\bibfnamefont {D.~T.}\ \bibnamefont
  {Adroja}}, \bibinfo {author} {\bibfnamefont {W.}~\bibnamefont {Kockelmann}},
  \bibinfo {author} {\bibfnamefont {A.~D.}\ \bibnamefont {Hillier}}, \bibinfo
  {author} {\bibfnamefont {J.~Y.}\ \bibnamefont {So}}, \bibinfo {author}
  {\bibfnamefont {K.~S.}\ \bibnamefont {Knight}}, and\ \bibinfo {author}
  {\bibfnamefont {B.~D.}\ \bibnamefont {Rainford}},\ }\bibfield  {title}
  {\bibinfo {title} {Reduced moment magnetic ordering in a Kondo lattice
  compound: {${\mathrm{Ce}}_{8}{\mathrm{Pd}}_{24}\mathrm{Ga}$}},\ }\href
  {https://doi.org/10.1103/PhysRevB.67.134419} {\bibfield  {journal} {\bibinfo
  {journal} {Phys. Rev. B}\ }\textbf {\bibinfo {volume} {67}},\ \bibinfo
  {pages} {134419} (\bibinfo {year} {2003})}\BibitemShut {NoStop}%
\bibitem [{\citenamefont {Hafner}\ \emph {et~al.}(2019)\citenamefont {Hafner},
  \citenamefont {Rai}, \citenamefont {Banda}, \citenamefont {Kliemt},
  \citenamefont {Krellner}, \citenamefont {Sichelschmidt}, \citenamefont
  {Morosan}, \citenamefont {Geibel},\ and\ \citenamefont
  {Brando}}]{Hafner2019}%
  \BibitemOpen
  \bibfield  {author} {\bibinfo {author} {\bibfnamefont {D.}~\bibnamefont
  {Hafner}}, \bibinfo {author} {\bibfnamefont {B.~K.}\ \bibnamefont {Rai}},
  \bibinfo {author} {\bibfnamefont {J.}~\bibnamefont {Banda}}, \bibinfo
  {author} {\bibfnamefont {K.}~\bibnamefont {Kliemt}}, \bibinfo {author}
  {\bibfnamefont {C.}~\bibnamefont {Krellner}}, \bibinfo {author}
  {\bibfnamefont {J.}~\bibnamefont {Sichelschmidt}}, \bibinfo {author}
  {\bibfnamefont {E.}~\bibnamefont {Morosan}}, \bibinfo {author} {\bibfnamefont
  {C.}~\bibnamefont {Geibel}}, and\ \bibinfo {author} {\bibfnamefont
  {M.}~\bibnamefont {Brando}},\ }\bibfield  {title} {\bibinfo {title}
  {Kondo-lattice ferromagnets and their peculiar order along the magnetically
  hard axis determined by the crystalline electric field},\ }\href
  {https://doi.org/10.1103/PhysRevB.99.201109} {\bibfield  {journal} {\bibinfo
  {journal} {Phys. Rev. B}\ }\textbf {\bibinfo {volume} {99}},\ \bibinfo
  {pages} {201109(R)} (\bibinfo {year} {2019})}\BibitemShut {NoStop}%
\bibitem [{\citenamefont {Kittler}\ \emph {et~al.}(2013)\citenamefont
  {Kittler}, \citenamefont {Fritsch}, \citenamefont {Weber}, \citenamefont
  {Fischer}, \citenamefont {Lamago}, \citenamefont {Andr\'e},\ and\
  \citenamefont {v.~L\"ohneysen}}]{Kittler2013}%
  \BibitemOpen
  \bibfield  {author} {\bibinfo {author} {\bibfnamefont {W.}~\bibnamefont
  {Kittler}}, \bibinfo {author} {\bibfnamefont {V.}~\bibnamefont {Fritsch}},
  \bibinfo {author} {\bibfnamefont {F.}~\bibnamefont {Weber}}, \bibinfo
  {author} {\bibfnamefont {G.}~\bibnamefont {Fischer}}, \bibinfo {author}
  {\bibfnamefont {D.}~\bibnamefont {Lamago}}, \bibinfo {author} {\bibfnamefont
  {G.}~\bibnamefont {Andr\'e}}, and\ \bibinfo {author} {\bibfnamefont
  {H.}~\bibnamefont {v.~L\"ohneysen}},\ }\bibfield  {title} {\bibinfo {title}
  {Suppression of ferromagnetism of {CeTiGe${}_{3}$ by V} substitution},\
  }\href {https://doi.org/10.1103/PhysRevB.88.165123} {\bibfield  {journal}
  {\bibinfo  {journal} {Phys. Rev. B}\ }\textbf {\bibinfo {volume} {88}},\
  \bibinfo {pages} {165123} (\bibinfo {year} {2013})}\BibitemShut {NoStop}%
\bibitem [{\citenamefont {Inamdar}\ \emph {et~al.}(2014)\citenamefont
  {Inamdar}, \citenamefont {Thamizhavel},\ and\ \citenamefont
  {Dhar}}]{Inamdar2014}%
  \BibitemOpen
  \bibfield  {author} {\bibinfo {author} {\bibfnamefont {M.}~\bibnamefont
  {Inamdar}}, \bibinfo {author} {\bibfnamefont {A.}~\bibnamefont
  {Thamizhavel}}, and\ \bibinfo {author} {\bibfnamefont {S.~K.}\ \bibnamefont
  {Dhar}},\ }\bibfield  {title} {\bibinfo {title} {Anisotropic magnetic
  behavior of single crystalline {CeTiGe$_3$} and {CeVGe$_3$}},\ }\href
  {https://doi.org/10.1088/0953-8984/26/32/326003} {\bibfield  {journal}
  {\bibinfo  {journal} {J. Phys. Condens. Matter}\ }\textbf {\bibinfo {volume}
  {26}},\ \bibinfo {pages} {326003} (\bibinfo {year} {2014})}\BibitemShut
  {NoStop}%
\bibitem [{\citenamefont {Majumder}\ \emph {et~al.}(2019)\citenamefont
  {Majumder}, \citenamefont {Kittler}, \citenamefont {Fritsch}, \citenamefont
  {L\"ohneysen}, \citenamefont {Yasuoka},\ and\ \citenamefont
  {Baenitz}}]{Majumder2019}%
  \BibitemOpen
  \bibfield  {author} {\bibinfo {author} {\bibfnamefont {M.}~\bibnamefont
  {Majumder}}, \bibinfo {author} {\bibfnamefont {W.}~\bibnamefont {Kittler}},
  \bibinfo {author} {\bibfnamefont {V.}~\bibnamefont {Fritsch}}, \bibinfo
  {author} {\bibfnamefont {H.~v.}\ \bibnamefont {L\"ohneysen}}, \bibinfo
  {author} {\bibfnamefont {H.}~\bibnamefont {Yasuoka}}, and\ \bibinfo {author}
  {\bibfnamefont {M.}~\bibnamefont {Baenitz}},\ }\bibfield  {title} {\bibinfo
  {title} {Competing magnetic correlations across the ferromagnetic quantum
  critical point in the {Kondo system
  ${\mathrm{CeTi}}_{1\ensuremath{-}x}{\mathrm{V}}_{x}{\mathrm{Ge}}_{3}$:
  $^{51}\mathrm{V}$ NMR} as a local probe},\ }\href
  {https://doi.org/10.1103/PhysRevB.100.134432} {\bibfield  {journal} {\bibinfo
   {journal} {Phys. Rev. B}\ }\textbf {\bibinfo {volume} {100}},\ \bibinfo
  {pages} {134432} (\bibinfo {year} {2019})}\BibitemShut {NoStop}%
\bibitem [{\citenamefont {Katoh}\ \emph {et~al.}(2009)\citenamefont {Katoh},
  \citenamefont {Nakagawa}, \citenamefont {Terui},\ and\ \citenamefont
  {Ochiai}}]{Katoh2009}%
  \BibitemOpen
  \bibfield  {author} {\bibinfo {author} {\bibfnamefont {K.}~\bibnamefont
  {Katoh}}, \bibinfo {author} {\bibfnamefont {S.}~\bibnamefont {Nakagawa}},
  \bibinfo {author} {\bibfnamefont {G.}~\bibnamefont {Terui}}, and\ \bibinfo
  {author} {\bibfnamefont {A.}~\bibnamefont {Ochiai}},\ }\bibfield  {title}
  {\bibinfo {title} {Magnetic and transport properties of single-crystal
  {YbPtGe}},\ }\href {https://doi.org/10.1143/JPSJ.78.104721} {\bibfield
  {journal} {\bibinfo  {journal} {J. Phys. Soc. Jpn.}\ }\textbf {\bibinfo
  {volume} {78}},\ \bibinfo {pages} {104721} (\bibinfo {year}
  {2009})}\BibitemShut {NoStop}%
\bibitem [{\citenamefont {Krellner}\ and\ \citenamefont
  {Geibel}(2012)}]{Krellner2012}%
  \BibitemOpen
  \bibfield  {author} {\bibinfo {author} {\bibfnamefont {C.}~\bibnamefont
  {Krellner}} and\ \bibinfo {author} {\bibfnamefont {C.}~\bibnamefont
  {Geibel}},\ }\bibfield  {title} {\bibinfo {title} {Magnetic anisotropy of
  {YbNi$_4$P$_2$}},\ }\href {https://doi.org/10.1088/1742-6596/391/1/012032}
  {\bibfield  {journal} {\bibinfo  {journal} {J. Phys.: Conf. Ser. 391 012032
  (2012)}\ }\textbf {\bibinfo {volume} {391}},\ \bibinfo {pages} {012032}
  (\bibinfo {year} {2012})}\BibitemShut {NoStop}%
\bibitem [{\citenamefont {Rai}\ \emph {et~al.}(2019)\citenamefont {Rai},
  \citenamefont {Stavinoha}, \citenamefont {Banda}, \citenamefont {Hafner},
  \citenamefont {Benavides}, \citenamefont {Sokolov}, \citenamefont {Chan},
  \citenamefont {Brando}, \citenamefont {Huang},\ and\ \citenamefont
  {Morosan}}]{Rai2019}%
  \BibitemOpen
  \bibfield  {author} {\bibinfo {author} {\bibfnamefont {B.~K.}\ \bibnamefont
  {Rai}}, \bibinfo {author} {\bibfnamefont {M.}~\bibnamefont {Stavinoha}},
  \bibinfo {author} {\bibfnamefont {J.}~\bibnamefont {Banda}}, \bibinfo
  {author} {\bibfnamefont {D.}~\bibnamefont {Hafner}}, \bibinfo {author}
  {\bibfnamefont {K.~A.}\ \bibnamefont {Benavides}}, \bibinfo {author}
  {\bibfnamefont {D.~A.}\ \bibnamefont {Sokolov}}, \bibinfo {author}
  {\bibfnamefont {J.~Y.}\ \bibnamefont {Chan}}, \bibinfo {author}
  {\bibfnamefont {M.}~\bibnamefont {Brando}}, \bibinfo {author} {\bibfnamefont
  {C.-L.}\ \bibnamefont {Huang}}, and\ \bibinfo {author} {\bibfnamefont
  {E.}~\bibnamefont {Morosan}},\ }\bibfield  {title} {\bibinfo {title}
  {Ferromagnetic ordering along the hard axis in the Kondo lattice
  ${\mathrm{YbIr}}_{3}{\mathrm{Ge}}_{7}$},\ }\href
  {https://doi.org/10.1103/PhysRevB.99.121109} {\bibfield  {journal} {\bibinfo
  {journal} {Phys. Rev. B}\ }\textbf {\bibinfo {volume} {99}},\ \bibinfo
  {pages} {121109(R)} (\bibinfo {year} {2019})}\BibitemShut {NoStop}%
\bibitem [{\citenamefont {Gruner}\ and\ \citenamefont
  {Zawadowski}(1974)}]{Gruner1974}%
  \BibitemOpen
  \bibfield  {author} {\bibinfo {author} {\bibfnamefont {G.}~\bibnamefont
  {Gruner}} and\ \bibinfo {author} {\bibfnamefont {A.}~\bibnamefont
  {Zawadowski}},\ }\bibfield  {title} {\bibinfo {title} {Magnetic impurities in
  non-magnetic metals},\ }\href {https://doi.org/10.1088/0034-4885/37/12/001}
  {\bibfield  {journal} {\bibinfo  {journal} {Rep. Prog. Phys.}\ }\textbf
  {\bibinfo {volume} {37}},\ \bibinfo {pages} {1497} (\bibinfo {year}
  {1974})}\BibitemShut {NoStop}%
\bibitem [{\citenamefont {Krishna-murthy}\ \emph {et~al.}(1975)\citenamefont
  {Krishna-murthy}, \citenamefont {Wilson},\ and\ \citenamefont
  {Wilkins}}]{Krishna-murthy1975}%
  \BibitemOpen
  \bibfield  {author} {\bibinfo {author} {\bibfnamefont {H.~R.}\ \bibnamefont
  {Krishna-murthy}}, \bibinfo {author} {\bibfnamefont {K.~G.}\ \bibnamefont
  {Wilson}}, and\ \bibinfo {author} {\bibfnamefont {J.~W.}\ \bibnamefont
  {Wilkins}},\ }\bibfield  {title} {\bibinfo {title} {Temperature-dependent
  susceptibility of the symmetric {A}nderson model: Connection to the {K}ondo
  model},\ }\href {https://doi.org/10.1103/PhysRevLett.35.1101} {\bibfield
  {journal} {\bibinfo  {journal} {Phys. Rev. Lett.}\ }\textbf {\bibinfo
  {volume} {35}},\ \bibinfo {pages} {1101} (\bibinfo {year}
  {1975})}\BibitemShut {NoStop}%
\bibitem [{MER()}]{MERLINdata}%
  \BibitemOpen
  \href@noop {} {}\bibinfo {note} {{M. Smidman et al; (2018): STFC ISIS Neutron
  and Muon Source, \url{https://doi.org/10.5286/ISIS.E.RB1820492}}}\BibitemShut
  {NoStop}%
\bibitem [{OSI()}]{OSIRISdata}%
  \BibitemOpen
  \href@noop {} {}\bibinfo {note} {{M. Smidman et al; (2018): STFC ISIS Neutron
  and Muon Source, \url{https://doi.org/10.5286/ISIS.E.RB1820611}}}\BibitemShut
  {NoStop}%
\bibitem [{\citenamefont {Bauer}(1991)}]{Bauer1991}%
  \BibitemOpen
  \bibfield  {author} {\bibinfo {author} {\bibfnamefont {E.}~\bibnamefont
  {Bauer}},\ }\bibfield  {title} {\bibinfo {title} {Anomalous properties of
  {Ce-Cu- and Yb-Cu-}based compounds},\ }\href
  {https://doi.org/10.1080/00018739100101512} {\bibfield  {journal} {\bibinfo
  {journal} {Advances in Physics}\ }\textbf {\bibinfo {volume} {40}},\ \bibinfo
  {pages} {417} (\bibinfo {year} {1991})}\BibitemShut {NoStop}%
\bibitem [{\citenamefont {Hattori}(2010)}]{Hattori2010}%
  \BibitemOpen
  \bibfield  {author} {\bibinfo {author} {\bibfnamefont {K.}~\bibnamefont
  {Hattori}},\ }\bibfield  {title} {\bibinfo {title} {Meta-orbital transition
  in heavy-fermion systems: {A}nalysis by dynamical mean field theory and
  self-consistent renormalization theory of orbital fluctuations},\ }\href
  {https://doi.org/10.1143/JPSJ.79.114717} {\bibfield  {journal} {\bibinfo
  {journal} {J. Phys. Soc. Jpn.}\ }\textbf {\bibinfo {volume} {79}},\ \bibinfo
  {pages} {114717} (\bibinfo {year} {2010})}\BibitemShut {NoStop}%
\bibitem [{\citenamefont {Pourovskii}\ \emph {et~al.}(2014)\citenamefont
  {Pourovskii}, \citenamefont {Hansmann}, \citenamefont {Ferrero},\ and\
  \citenamefont {Georges}}]{Pourovskii2014}%
  \BibitemOpen
  \bibfield  {author} {\bibinfo {author} {\bibfnamefont {L.~V.}\ \bibnamefont
  {Pourovskii}}, \bibinfo {author} {\bibfnamefont {P.}~\bibnamefont
  {Hansmann}}, \bibinfo {author} {\bibfnamefont {M.}~\bibnamefont {Ferrero}},
  and\ \bibinfo {author} {\bibfnamefont {A.}~\bibnamefont {Georges}},\
  }\bibfield  {title} {\bibinfo {title} {Theoretical prediction and
  spectroscopic fingerprints of an orbital transition in
  {${\mathrm{CeCu}}_{2}{\mathrm{Si}}_{2}$}},\ }\href
  {https://doi.org/10.1103/PhysRevLett.112.106407} {\bibfield  {journal}
  {\bibinfo  {journal} {Phys. Rev. Lett.}\ }\textbf {\bibinfo {volume} {112}},\
  \bibinfo {pages} {106407} (\bibinfo {year} {2014})}\BibitemShut {NoStop}%
\bibitem [{\citenamefont {Amorese}\ \emph {et~al.}(2020)\citenamefont
  {Amorese}, \citenamefont {Marino}, \citenamefont {Sundermann}, \citenamefont
  {Chen}, \citenamefont {Hu}, \citenamefont {Willers}, \citenamefont
  {Choukani}, \citenamefont {Ohresser}, \citenamefont {Herrero-Martin},
  \citenamefont {Agrestini}, \citenamefont {Chen}, \citenamefont {Lin},
  \citenamefont {Haverkort}, \citenamefont {Seiro}, \citenamefont {Geibel},
  \citenamefont {Steglich}, \citenamefont {Tjeng}, \citenamefont {Zwicknagl},\
  and\ \citenamefont {Severing}}]{Amorese2020}%
  \BibitemOpen
  \bibfield  {author} {\bibinfo {author} {\bibfnamefont {A.}~\bibnamefont
  {Amorese}}, \bibinfo {author} {\bibfnamefont {A.}~\bibnamefont {Marino}},
  \bibinfo {author} {\bibfnamefont {M.}~\bibnamefont {Sundermann}}, \bibinfo
  {author} {\bibfnamefont {K.}~\bibnamefont {Chen}}, \bibinfo {author}
  {\bibfnamefont {Z.}~\bibnamefont {Hu}}, \bibinfo {author} {\bibfnamefont
  {T.}~\bibnamefont {Willers}}, \bibinfo {author} {\bibfnamefont
  {F.}~\bibnamefont {Choueikani}}, \bibinfo {author} {\bibfnamefont
  {P.}~\bibnamefont {Ohresser}}, \bibinfo {author} {\bibfnamefont
  {J.}~\bibnamefont {Herrero-Martin}}, \bibinfo {author} {\bibfnamefont
  {S.}~\bibnamefont {Agrestini}}, \bibinfo {author} {\bibfnamefont {C.-T.}\
  \bibnamefont {Chen}}, \bibinfo {author} {\bibfnamefont {H.-J.}\ \bibnamefont
  {Lin}}, \bibinfo {author} {\bibfnamefont {M.~W.}\ \bibnamefont {Haverkort}},
  \bibinfo {author} {\bibfnamefont {S.}~\bibnamefont {Seiro}}, \bibinfo
  {author} {\bibfnamefont {C.}~\bibnamefont {Geibel}}, \bibinfo {author}
  {\bibfnamefont {F.}~\bibnamefont {Steglich}}, \bibinfo {author}
  {\bibfnamefont {L.~H.}\ \bibnamefont {Tjeng}}, \bibinfo {author}
  {\bibfnamefont {G.}~\bibnamefont {Zwicknagl}}, and\ \bibinfo {author}
  {\bibfnamefont {A.}~\bibnamefont {Severing}},\ }\bibfield  {title} {\bibinfo
  {title} {Possible multiorbital ground state in
  {${\mathrm{CeCu}}_{2}{\mathrm{Si}}_{2}$}},\ }\href
  {https://doi.org/10.1103/PhysRevB.102.245146} {\bibfield  {journal} {\bibinfo
   {journal} {Phys. Rev. B}\ }\textbf {\bibinfo {volume} {102}},\ \bibinfo
  {pages} {245146} (\bibinfo {year} {2020})}\BibitemShut {NoStop}%
\bibitem [{\citenamefont {Willers}\ \emph {et~al.}(2010)\citenamefont
  {Willers}, \citenamefont {Hu}, \citenamefont {Hollmann}, \citenamefont
  {K\"orner}, \citenamefont {Gegner}, \citenamefont {Burnus}, \citenamefont
  {Fujiwara}, \citenamefont {Tanaka}, \citenamefont {Schmitz}, \citenamefont
  {Hsieh}, \citenamefont {Lin}, \citenamefont {Chen}, \citenamefont {Bauer},
  \citenamefont {Sarrao}, \citenamefont {Goremychkin}, \citenamefont {Koza},
  \citenamefont {Tjeng},\ and\ \citenamefont {Severing}}]{Willers2010}%
  \BibitemOpen
  \bibfield  {author} {\bibinfo {author} {\bibfnamefont {T.}~\bibnamefont
  {Willers}}, \bibinfo {author} {\bibfnamefont {Z.}~\bibnamefont {Hu}},
  \bibinfo {author} {\bibfnamefont {N.}~\bibnamefont {Hollmann}}, \bibinfo
  {author} {\bibfnamefont {P.~O.}\ \bibnamefont {K\"orner}}, \bibinfo {author}
  {\bibfnamefont {J.}~\bibnamefont {Gegner}}, \bibinfo {author} {\bibfnamefont
  {T.}~\bibnamefont {Burnus}}, \bibinfo {author} {\bibfnamefont
  {H.}~\bibnamefont {Fujiwara}}, \bibinfo {author} {\bibfnamefont
  {A.}~\bibnamefont {Tanaka}}, \bibinfo {author} {\bibfnamefont
  {D.}~\bibnamefont {Schmitz}}, \bibinfo {author} {\bibfnamefont {H.~H.}\
  \bibnamefont {Hsieh}}, \bibinfo {author} {\bibfnamefont {H.-J.}\ \bibnamefont
  {Lin}}, \bibinfo {author} {\bibfnamefont {C.~T.}\ \bibnamefont {Chen}},
  \bibinfo {author} {\bibfnamefont {E.~D.}\ \bibnamefont {Bauer}}, \bibinfo
  {author} {\bibfnamefont {J.~L.}\ \bibnamefont {Sarrao}}, \bibinfo {author}
  {\bibfnamefont {E.}~\bibnamefont {Goremychkin}}, \bibinfo {author}
  {\bibfnamefont {M.}~\bibnamefont {Koza}}, \bibinfo {author} {\bibfnamefont
  {L.~H.}\ \bibnamefont {Tjeng}}, and\ \bibinfo {author} {\bibfnamefont
  {A.}~\bibnamefont {Severing}},\ }\bibfield  {title} {\bibinfo {title}
  {Crystal-field and {K}ondo-scale investigations of
  {$\text{Ce}M{\text{In}}_{5}$ ($M=\text{Co}$, Ir, and Rh): A} combined x-ray
  absorption and inelastic neutron scattering study},\ }\href
  {https://doi.org/10.1103/PhysRevB.81.195114} {\bibfield  {journal} {\bibinfo
  {journal} {Phys. Rev. B}\ }\textbf {\bibinfo {volume} {81}},\ \bibinfo
  {pages} {195114} (\bibinfo {year} {2010})}\BibitemShut {NoStop}%
\bibitem [{\citenamefont {Willers}\ \emph {et~al.}(2015)\citenamefont
  {Willers}, \citenamefont {Strigari}, \citenamefont {Hu}, \citenamefont
  {Sessi}, \citenamefont {Brookes}, \citenamefont {Bauer}, \citenamefont
  {Sarrao}, \citenamefont {Thompson}, \citenamefont {Tanaka}, \citenamefont
  {Wirth}, \citenamefont {Tjeng},\ and\ \citenamefont
  {Severing}}]{Willers2015}%
  \BibitemOpen
  \bibfield  {author} {\bibinfo {author} {\bibfnamefont {T.}~\bibnamefont
  {Willers}}, \bibinfo {author} {\bibfnamefont {F.}~\bibnamefont {Strigari}},
  \bibinfo {author} {\bibfnamefont {Z.}~\bibnamefont {Hu}}, \bibinfo {author}
  {\bibfnamefont {V.}~\bibnamefont {Sessi}}, \bibinfo {author} {\bibfnamefont
  {N.~B.}\ \bibnamefont {Brookes}}, \bibinfo {author} {\bibfnamefont {E.~D.}\
  \bibnamefont {Bauer}}, \bibinfo {author} {\bibfnamefont {J.~L.}\ \bibnamefont
  {Sarrao}}, \bibinfo {author} {\bibfnamefont {J.~D.}\ \bibnamefont
  {Thompson}}, \bibinfo {author} {\bibfnamefont {A.}~\bibnamefont {Tanaka}},
  \bibinfo {author} {\bibfnamefont {S.}~\bibnamefont {Wirth}}, \bibinfo
  {author} {\bibfnamefont {L.~H.}\ \bibnamefont {Tjeng}}, and\ \bibinfo
  {author} {\bibfnamefont {A.}~\bibnamefont {Severing}},\ }\bibfield  {title}
  {\bibinfo {title} {Correlation between ground state and orbital anisotropy in
  heavy fermion materials},\ }\href {https://doi.org/10.1073/pnas.1415657112}
  {\bibfield  {journal} {\bibinfo  {journal} {Proc. Natl. Acad. Sci. U.S.A}\
  }\textbf {\bibinfo {volume} {112}},\ \bibinfo {pages} {2384} (\bibinfo {year}
  {2015})}\BibitemShut {NoStop}%
\bibitem [{\citenamefont {Shim}\ \emph {et~al.}(2007)\citenamefont {Shim},
  \citenamefont {Haule},\ and\ \citenamefont {Kotliar}}]{Shim2007}%
  \BibitemOpen
  \bibfield  {author} {\bibinfo {author} {\bibfnamefont {J.~H.}\ \bibnamefont
  {Shim}}, \bibinfo {author} {\bibfnamefont {K.}~\bibnamefont {Haule}}, and\
  \bibinfo {author} {\bibfnamefont {G.}~\bibnamefont {Kotliar}},\ }\bibfield
  {title} {\bibinfo {title} {Modeling the localized-to-itinerant electronic
  transition in the heavy fermion system {CeIrIn$_5$}},\ }\href
  {https://doi.org/10.1126/science.1149064} {\bibfield  {journal} {\bibinfo
  {journal} {Science}\ }\textbf {\bibinfo {volume} {318}},\ \bibinfo {pages}
  {1615} (\bibinfo {year} {2007})}\BibitemShut {NoStop}%
\end{thebibliography}
\end{document}